\documentclass[draft]{agujournal2019}
\usepackage{url} 
\usepackage{lineno}
\usepackage[inline]{trackchanges} 
\usepackage{soul}
\usepackage{bbold}
\usepackage{mathtools}
\draftfalse

\journalname{Enter journal name here}

\begin{document}

%
%


\title{Subsurface ocean salinity and dissipation rate inferred from Enceladus ice shell morphology}

%
%




\authors{Wanying Kang\affil{1}, Yixiao Zhang\affil{1}}


\affiliation{1}{Earth, Atmospheric and Planetary Science Department, Massachusetts Institute of Technology, Cambridge, MA 02139, USA}




\correspondingauthor{Wanying Kang}{wanying@mit.edu}



\begin{keypoints}
\item Equatorward heat convergence by ocean circulation on Enceladus is enhanced by strong tidal mixing and extreme salinity values.
\item The sustainability of Enceladus' strong ice thickness sets an upper bound on the heat convergence.
\item This upper bound can be used to constrain Enceladus ocean salinity and tidal mixing using our scaling laws. 
\end{keypoints}

%
%

%
%


\begin{abstract}
  The habitability of Enceladus’ subsurface ocean and the detectability of potential biosignatures depend on efficient ocean circulation and suitable ocean conditions. Directly probing the ocean is challenging because it lies beneath a thick ice shell; however, the ice thickness distribution is relatively well constrained and provides indirect insight into the underlying ocean dynamics. This study investigates how ocean circulation and the associated heat transport depend on ocean salinity and tide-induced vertical mixing using scaling analysis, supported by numerical simulations. We find that ocean circulation and equatorward heat convergence are stronger under extremely high or low salinity conditions than under intermediate salinity, and both increase with tidal mixing rates. Because the poleward thinning of Enceladus’ ice shell cannot be maintained in the presence of strong equatorward ocean heat transport, these results place constraints on the ocean salinity, diffusivity, circulation timescale, and ocean dissipation rate. Energetic analysis further shows that Enceladus’ ocean behaves like an extremely efficient heat pump (inefficient heat engine), potentially transporting up to $1000$ times more heat across latitudes than the energy dissipated within the ocean itself, thereby placing strong constraints on the ocean’s energy dissipation rate.
\end{abstract}


\section*{Plain Language Summary}

The habitability of Enceladus' hidden ocean depends heavily on how effectively the ocean circulates. Though delving into the secrets of this ocean shrouded beneath a thick layer of ice presents a challenge, the observed ice thickness profile may provide valuable insights into the ocean below. This study delves into how ocean currents and the transport of heat vary based on factors such as the ocean's saltiness and the mixing caused by tidal forces. Through scaling analysis and computer simulations, we discover that the ocean currents and heat transport intensify when the ocean's saltiness reaches extremes or when tides vigorously mix things up. However, to sustain the strong poleward thinning trend of Enceladus ice shell, the heat convergence toward the equator cannot be arbitrarily strong. This constraint can be used to infer the salinity, the tidal mixing rate, as well as the circulation rate in the ocean deep below the surface.

%
%

%


%
%
%
%

\section{Introduction}
Many icy satellites in our solar system have been found to contain a global subsurface ocean \cite{Carr-Belton-Chapman-et-al-1998:evidence, Kivelson-Khurana-Russell-et-al-2000:galileo, Pappalardo-Belton-Breneman-et-al-1999:does, Thomas-Tajeddine-Tiscareno-et-al-2016:enceladus, Khurana-Kivelson-Stevenson-et-al-1998:induced, Kivelson-Khurana-Stevenson-et-al-1999:europa, Zimmer-Khurana-Kivelson-2000:subsurface, Hand-Chyba-2007:empirical}, which is potentially habitable \cite{Glein-Waite-2020:carbonate, Taubner-Pappenreiter-Zwicker-et-al-2018:biological, Chyba-2000:energy, Russell-Murray-Hand-2017:possible}. Habitability and its detectability depend on the presence of an ocean circulation capable of continuously replenishing nutrients and transporting potential biosignatures to the surface \cite{Cockell-Simons-Castillo-Rogez-et-al-2023:sustained}. The ocean circulation is simultaneously forced by heat flux from the silicate core \cite{Bire-Kang-Ramadhan-et-al-2022:exploring, Soderlund-Schmidt-Wicht-et-al-2014:ocean, Ashkenazy-Tziperman-2021:dynamic, Kang-2023:modulation}, heat/salinity exchanges with the ice shell \cite{Zhu-Manucharyan-Thompson-et-al-2017:influence, Kang-Jansen-2022:icy, Kang-2022:different, Zhang-Kang-Marshall-2024:ocean} and tides \cite{Rovira-Navarro-Rieutord-Gerkema-et-al-2019:do,Rekier-Trinh-Triana-et-al-2019:internal,Hay-Matsuyama-2019:nonlinear}. Its magnitude and direction is thus determined by the partitioning of dissipation between the ice shell, the silicate core and the ocean as well as the mean salinity in the ocean through the equation of state \cite{Melosh-Ekholm-Showman-et-al-2004:temperature, Zeng-Jansen-2021:ocean,Kang-Mittal-Bire-et-al-2022:how}. Covered by a global ice shell tens of kilometers thick, these properties and forcing conditions of the subsurface oceans are difficult to observe.

On the other hand, the ice shell thickness variations can potentially be measured. Combining the gravity anomalies data and surface topography data, Enceladus's ice shell is revealed to present a strong poleward thinning trend \cite{Iess-Stevenson-Parisi-et-al-2014:gravity, Beuthe-Rivoldini-Trinh-2016:enceladuss, Hemingway-Iess-Tadjeddine-et-al-2018:interior, Hemingway-Mittal-2019:enceladuss, Schenk-McKinnon-2024:new, Park-Mastrodemos-Jacobson-et-al-2024:global}. As sketched in Fig.~\ref{fig:H-q}a, under the thick equatorial ice shell, water tends to be saltier and colder than the polar water. The temperature variability has to do with water's freezing point being suppressed by high pressure under thick ice, and the salinity variability is induced by freezing over the equator and melting over the poles, needed to sustain the poleward thinning trend against ice flow driven by the thickness gradients \cite{Zhu-Manucharyan-Thompson-et-al-2017:influence, Kang-Mittal-Bire-et-al-2022:how}. The resultant meridional temperature and salinity gradients can drive ocean circulation that transport heat from the poles, where the water is warm, toward the equator, where the water is cold.
If the strong equator-to-pole ice thickness variations are to be sustained, the ocean heat transport cannot be arbitrarily strong. This would allow us to put constraints on the aforementioned factors that can influence the ocean heat transport, including the partition of heat production between the ice shell and the silicate core, the strength of ocean tidal dissipation and the ocean salinity.

This work derives scaling laws for meridional heat transport by ocean circulation that is driven by heat and salinity forcings from the ice shell (section~\ref{sec:scaling}), and compares the analytical results against numerical simulations (section~\ref{sec:numerical}). In section.~\ref{sec:obs-implications}, we use the observed ice thickness profile for Enceladus to derive an upper bound of ocean heat transport, which is then converted into a constraint on the ocean salinity and ocean tidal dissipation rate.


\section{Temperature and salinity forcings in the system.}
\begin{figure}[hpt!]
\centering \includegraphics[width=\textwidth, page=5]{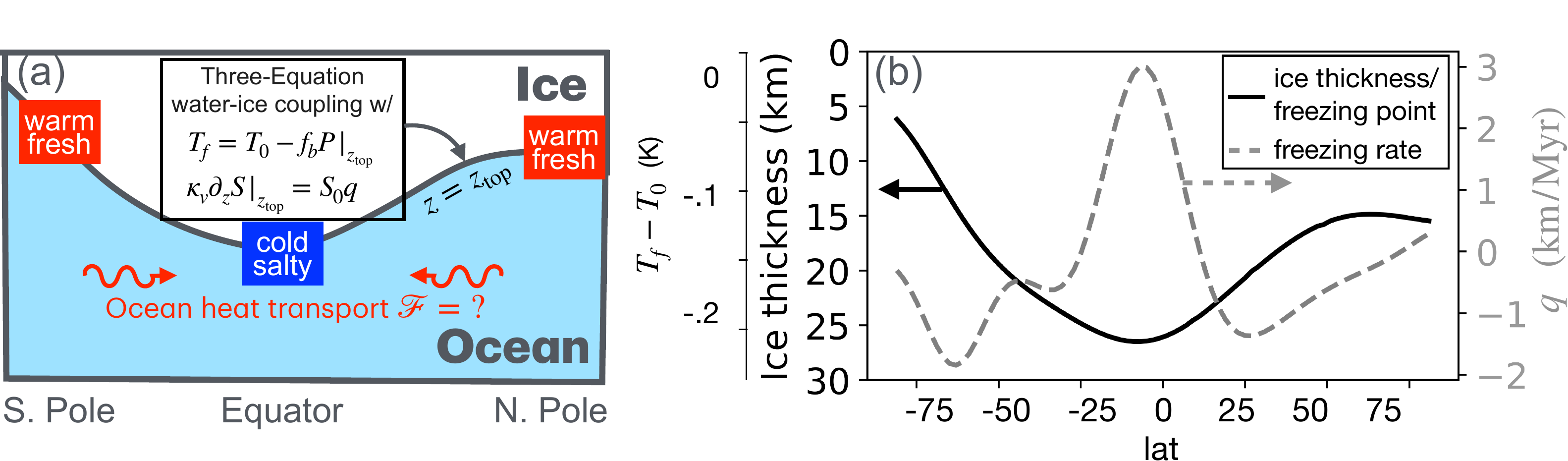}
\caption{Temperature and salinity forcings exerted on Enceladus ocean. Panel (a) presents the domain geometry and boundary condition at the water-ice interface. Panel (b) presents profiles of Enceladus ice thickness profile $H_i$ (solid) and the imposed freezing rate $q$. $q$ is set such that it balances out the ice flow (SI Text S2). Since water's freezing point $T_f$ depends almost linearly with $H_i$, a second y-axis is added to show the freezing point at water-ice interface. This study aims to estimate the ocean heat transport, $\mathcal{F}$, across a range of mean salinities ($S_0$) and vertical diffusivities ($\kappa_v$) to determine which combinations enable equatorial ice to freeze so that the ice thickness variations can be sustained against ice flow.}
\label{fig:H-q}
\end{figure}

We consider the ocean circulation driven by the aforementioned meridional temperature and salinity gradients along the water-ice interface, assuming zero bottom heating. Under Enceladus-like configuration, the equatorial side of the domain will be cold and salty, and the polar side of the domain will be warm and fresh (Fig.~\ref{fig:H-q}a).
The equator-to-pole temperature contrast $\Delta T$ at the water–ice interface arises from pressure-induced freezing suppression: the thicker equatorial ice shell exerts higher pressure, lowering the local melting point relative to the poles where the ice is thinner. 
Knowing the ice thickness gradient $\Delta H_i$, the under-ice temperature variation $\Delta T$ can be calculated
\begin{equation}
  \label{eq:DeltaT}
  \Delta T=f_b\rho_ig\Delta H_i,
\end{equation}
where $\rho_i$ denotes ice density, and $f_b$ denotes the sensitivity of freezing point to pressure changes, and $\Delta H_i$ denotes the equator-to-pole ice thickness variation on Enceladus.

The equator-to-pole salinity contrast $\Delta S$ arises from the freezing at the equator and melting at the poles, which is necessary to sustain the ice thickness variations against ice flow. The meridional ice transport is driven by pressure force induced by the higher ice surface at the equator in the hydrostatic state. Using a thin shell model, \citeA{Ashkenazy-Sayag-Tziperman-2018:dynamics} provided formula for the meridional ice transport, and the freezing rate $q$ (unit: m/s) needs to be equal to the divergence of this transport in order to prevent the ice shell geometry from changing. In our scaling analysis and numerical simulations, we set the salinity flux at the ocean top to
\begin{equation}
  \label{eq:salinity-flux}
  \left. \mathcal{F}_{S,v}\right|_{\rm top}=S_0q,
\end{equation}
where $S_0$ denotes the mean ocean salinity.

Adopting the ice thickness profile proposed by \citeA{Hemingway-Mittal-2019:enceladuss}, we calculate the corresponding salinity flux and under-ice temperature profiles for Enceladus, shown in Fig.~\ref{fig:H-q}b. Because equatorial water is always saltier and colder than polar water in our configuration, we define the temperature and salinity contrasts, $\Delta T$ and $\Delta S$, to be strictly positive for clarity. 
Our goal is to estimate the ocean heat transport, $\mathcal{F}$, across a range of mean salinities ($S_0$) and vertical diffusivities ($\kappa_v$) to determine which combinations enable equatorial ice to freeze so that the ice thickness variations can be sustained against ice flow.

\section{Scaling laws for ocean heat transport.}
\label{sec:scaling}

In this section, we derive scaling laws for the meridional ocean heat transport, $\mathcal{F}$, driven by temperature gradients and salinity fluxes at the water-ice interface. 
Scaling laws for icy moon ocean heat transport, driven solely by meridional temperature gradients, have been derived in \citeA{Kang-2022:different} and \citeA{Zhang-Kang-Marshall-2024:ocean} following the baroclinic turbulence theory \cite{Held-Larichev-1996:scaling}. In the derivation, salinity gradients induced by freezing/melting and the topography effect (i.e., the fact that warm water and cold water under the ice are not located at the same depth) are both neglected, as they tend to play a less important role on large icy moons in an equilibrium state \cite{Kang-Jansen-2022:icy}. Subsequently, \citeA{Zhang-Kang-Marshall-2025:how} generalized the scaling laws to account for the topography effect, and the goal of this work is to further add salinity factor, so that it can be applied to Enceladus.

The poleward warming and freshening trend in Enceladus’ ocean, induced by the poleward thinning of its ice shell, drives an ocean circulation that transports heat equatorward and salinity poleward, both down their respective gradients. This transport occurs primarily through baroclinic eddies, which extract potential energy from the meridional density gradients as dense fluid is advected downward and light fluid upward through eddy motions. Despite the complexity of the eddy motions, their horizontal and vertical transports $\mathcal{F}_{X,h},\ \mathcal{F}_{X,v}$ of tracer $X$ ($X$ can be temperature $T$, salinity $S$, or buoyancy $b$) can be equivalently represented by advection by an overturning circulation $\Psi^\dagger$ (chapter 12 of \cite{Vallis-2006:atmospheric}),
\begin{equation}
  \label{eq:horizontal-vertical-fluxes}
  \mathcal{F}_{X,h}= \int\Psi^\dagger\partial_z X~dz\sim \Psi^\dagger \Delta_zX,\ \mathcal{F}_{X,v}= \int-\Psi^\dagger \partial_yX~dy\sim \Psi^\dagger \Delta_yX,
\end{equation}
where $y=a\theta$ stands for the latitudinal distance and $z$ stands for depth. Accounting for the spherical geometry, $\Psi^\dagger$ can be written as $\Psi^\dagger_02\cos\theta$, where $\Psi^\dagger_0$ denotes the 60N/S value. With a vertical tracer contrast $\Delta_z X$, the circulation $\Psi^\dagger$ transports $X$ meridionally, because $X$ in the upper and lower branches differ. Similarly, with a meridional tracer contrast $\Delta_y X$, the circulation transports $X$ vertically.

The circulation $\Psi^\dagger$ can take either sign. As shown by \cite{Kang-Mittal-Bire-et-al-2022:how}, if the mean ocean salinity is high, water’s thermal expansion coefficient is positive, making equatorial water denser than polar water. In this case, the sinking branch occurs at the equator, and we denote this direction as $\Psi^\dagger_0>0$. However, if the mean ocean salinity is low, water’s thermal expansion coefficient becomes negative, so the cold water beneath the equatorial ice is more buoyant than the warmer polar water. The circulation then reverses direction, corresponding to $\Psi^\dagger_0<0$.

For simplicity, we consider an idealized tracer distribution as sketched in Fig.~\ref{fig:scaling}G, in which the isothermal, isosaline, and isopycnal surfaces originating from the water-ice interface extend into the ocean interior along straight lines. In scenarios 2 and 3, where these isolines intersect the seafloor, $\partial_y X$ and $\partial_z X$ remain constant throughout the domain. In scenarios 1 and 4, where the isolines do not reach the seafloor, the bottom of the ocean is filled with the densest fluid in the system, with uniform $X$ (i.e., $\partial_y X = \partial_z X = 0$).

In order to estimate the meridional heat transport $\mathcal{F}_{T,h}$, we need 1) a scaling law for $\Psi^\dagger_0$ that determines the tracer transport efficiency by baroclinic eddies, given the buoyancy distribution, characterized by isopycnal slope $s$, equator-to-pole temperature contrast $\Delta T$ and salinity contrast $\Delta S$ at the water-ice interface, and 2) the conditions for $s$ and $\Delta S$ to be in the equilibrium state.

The scaling law for $\Psi_0^\dagger$ is obtained by generalizing the results of \citeA{Kang-2022:different} and \citeA{Zhang-Kang-Marshall-2024:ocean} to account for salinity-induced density anomalies in addition to temperature-induced ones, as well as the influence of a tilted water–ice interface \cite{Zhang-Kang-Marshall-2025:how}. Without repeating the full derivation presented in \citeA{Zhang-Kang-Marshall-2025:how} (briefly summarized in SI Text S1), we provide the resulting expression for the circulation $\Psi^\dagger_0$ as a function of $s$, $\Delta T$, $\Delta S$, and planetary parameters.
\begin{equation}
   \label{eq:Psi}
  \Psi^{\dagger}_0= 
  \begin{dcases}
    \pi a \rho_0 k^{5/2}s^{5/2}\left(\frac{\Delta b}{2a}\right)^{3/2}\beta^{-2} \xi^{-3/2} \left(1+s_{\rm top}/s\right)^{-3/2},& \rm{if } \Delta b>0\\
    - \pi a \rho_0 k^{5/2}s^{5/2}\left(\frac{\Delta b}{2a}\right)^{3/2}\beta^{-2}\xi^{-3/2}\left(2as/D\right)^{3/2},              & \rm{if } \Delta b<0
\end{dcases} 
\end{equation}

Here, the buoyancy contrast under the ice $\Delta b$ is set by the under-ice $T$, $S$ contrasts 
\begin{equation}
  \label{eq:buoyancy}
  \Delta b=g(\alpha_T\Delta T+\beta_S\Delta S),
\end{equation}
where $\alpha_T$ and $\beta_S$ are the thermal expansivity and saline contractivity.
When both coefficients are positive, the temperature- and salinity-induced density anomalies reinforce each other, making equatorial water denser than polar water. However, at low salinity and low pressure, conditions that can occur on small icy moons, $\alpha_T$ becomes negative, causing the temperature- and salinity-induced density anomalies to partially cancel each other. If the temperature-induced density anomaly dominates, $\Delta b$ can become negative, leading to a reversed circulation ($\Psi^\dagger_0<0$), as illustrated in the fresh-ocean scenario of Fig.~\ref{fig:scaling}G.

\begin{figure}[hpt!]
\includegraphics[width=\linewidth]{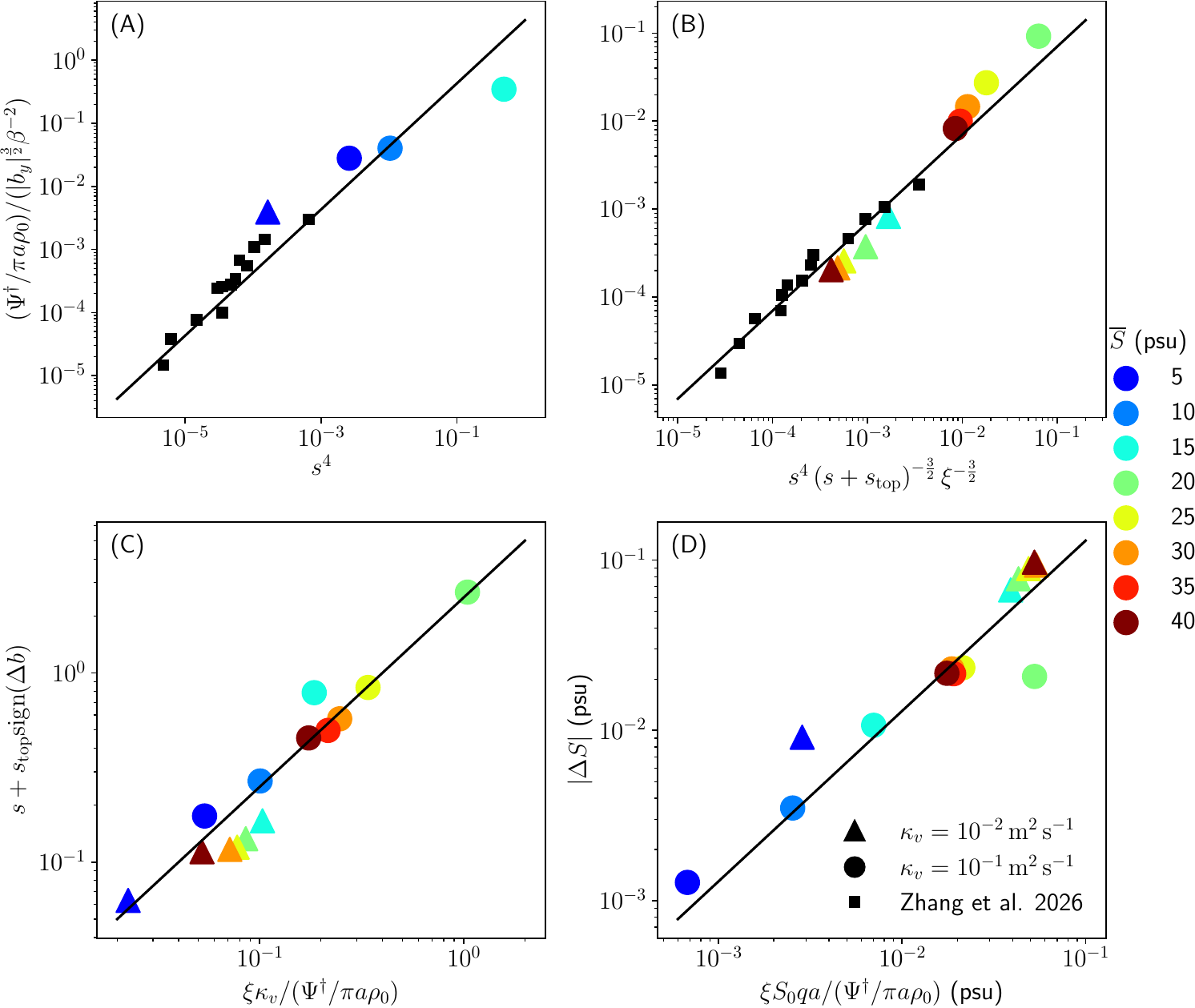}
\caption{Scaling laws for ocean heat transport tested using Oceananigans simulations. Panels (A) and (B) examine Eq.~\ref{eq:Psi} in the regimes of $\Delta b>0$ and $\Delta b<0$, respectively. Panel (C) tests the balance of vertical tracer transport (Eq.~\ref{eq:vertical-transport-balance}), and Panel (D) tests the balance of horizontal salinity transport (Eq.~\ref{eq:salinity-flux-balance}). Colored symbols represent the experiments conducted in this work: different colors correspond to different ocean salinities, while different shapes denote different ocean diffusivities. Black squares indicate experiments presented in \cite{Zhang-Kang-Marshall-2025:how}. 
}
\label{fig:scaling-scatter}
\end{figure}

\begin{figure}[hpt!]
\centering \includegraphics[width=\textwidth, page=6]{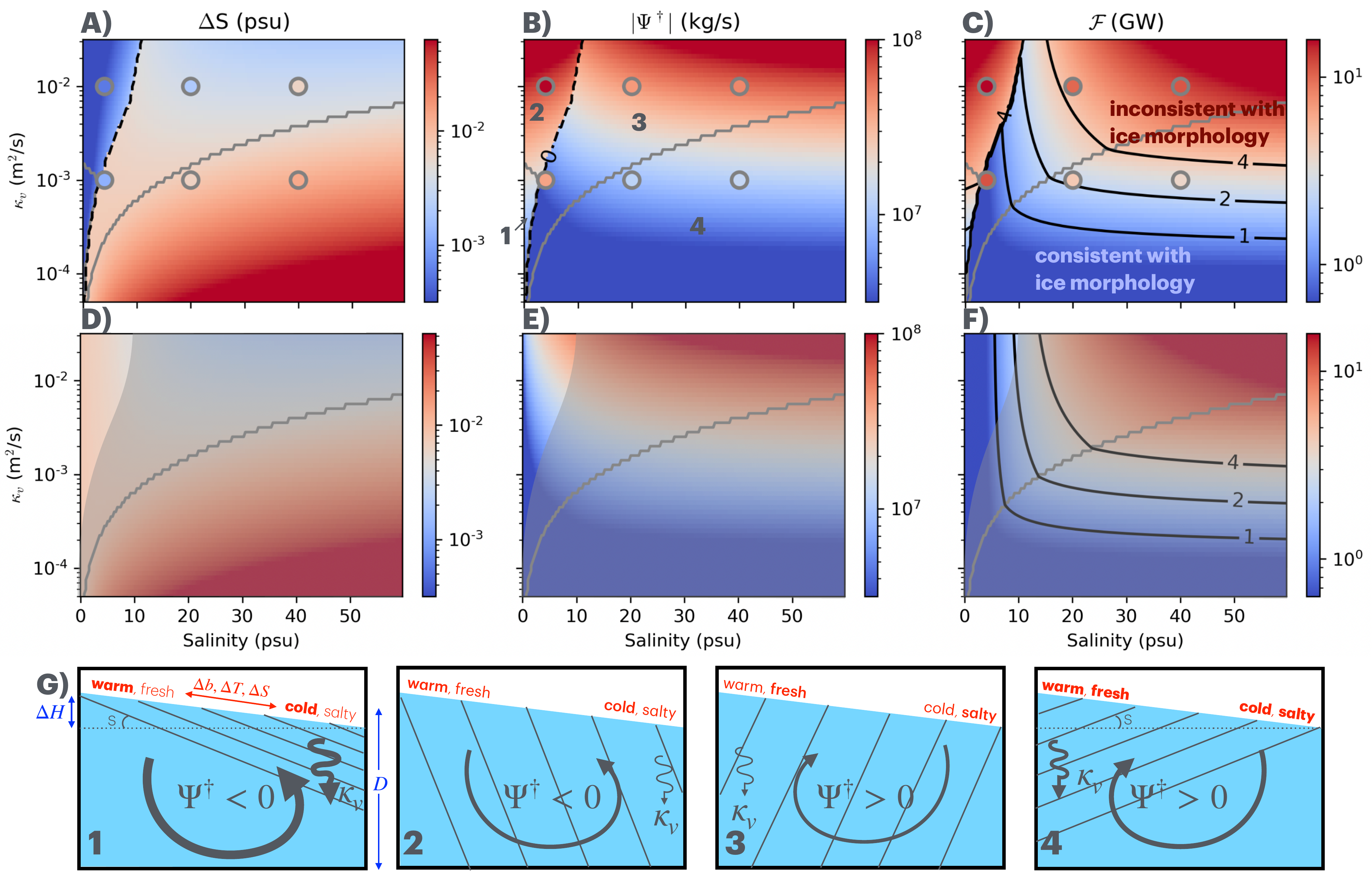}
\caption{Panels (A-C) present analytical predictions for the equator-to-pole salinity contrast $\Delta S$, the circulation strength $|\Psi^\dagger|$ and the equatorward heat transport $\mathcal{F}$. Separated by the grey curves, the upper part of the parameter space follows the D-limit scaling and the lower part follows the $\kappa_v$-limit scaling. Circulation reversal is denoted by the zero contour in the $\Psi^\dagger$ figure (panel B). The $\mathcal{F}=1,~2,~4$~GW (corresponding to 2,4,8~GW equatorward heat convergence) contours in panel (C) for reference. Besides the solution presented in panel (A-C), for the parameter regime with $\Psi^\dagger<0$, there is a set of different solution as presented in Panels (D-F). Shaded regions in panels (D-F) are identical to those in panel (A-C). The characteristics of ocean circulation (arrows) and isopycnals (contours) are sketched in panel (G) for the 4 scenarios with increasing ocean salinity from left to right. Their regimes are also marked on panel (B,E). Also denoted are the definitions of isopycnal slope $s$, the buoyancy contrast between equatorial and polar regions $\Delta b$.  
  }
\label{fig:scaling}
\end{figure}

To support the scaling given by Eq.~\ref{eq:Psi}, we conduct a set of numerical simulations using Oceananigans \cite{Ramadhan-Wagner-Hill-et-al-2020:oceananigans}. In these simulations, the overlying ice shell is prescribed to thin poleward following a sinusoidal profile:
\begin{equation}
z_{\rm top}=-10\rm{km}\cdot \cos(2\theta).
\end{equation}
The upper boundary conditions for temperature and salinity are set to
\begin{eqnarray}
\left.T\right|_{\rm top}-T_0&=&-0.5\rm{K}\cdot\cos(2\theta),\\
\left.\mathcal{F}_{S,v}\right|_{\rm top}&=&-\left.\kappa_v \partial_z S \right|_{\rm top}=10^{-9}\rm{m/s}\cdot S_0 \cos(2\theta).
\end{eqnarray}
Compared with what is typical for Enceladus, the temperature and salinity forcing amplitudes are enhanced by one order of magnitude to accelerate the convergence of the numerical simulations. We consider two values of the vertical diffusivity, $\kappa_v=0.1$ and $0.01\,\mathrm{m}^2\,\mathrm{s}^{-1}$, which are also considerably larger than the values expected for Enceladus. These enhanced diffusivities compensate for the stronger T/S forcing and prevent unrealistically small isopycnal slopes. We further consider a range of different mean ocean salinities $S_0$. Low-salinity experiments generate circulation that sinks at the poles ($\Psi^\dagger<0$), whereas high-salinity experiments generate circulation that sinks at the equator ($\Psi^\dagger>0$).
We then classify the experiments according to the direction of the circulation. Fig.~\ref{fig:scaling-scatter}A, B present the $\Psi$-$s$ scaling for the $\Psi^\dagger>0$ and $\Psi^\dagger<0$ cases, respectively. We also overlay the experiments from \citeA{Zhang-Kang-Marshall-2025:how} and compare them with the scaling laws given by Eq.~\ref{eq:Psi}. 
To measure $s$, we compute the slope of the median isopycnal, $s_\mathrm{m}$, by
tracking its latitudinal and vertical extent.
To estimate $\Psi$, we use the relation between horizontal heat transport and
the eddy overturning strength,
$\mathcal{F}_{T,h} \sim |\Psi_0^\dagger| \Delta T \, \xi^{-1}$, and define
the diagnosed
$|\Psi_0^\dagger| \equiv \xi\, \mathcal{F}_{T,h} /(0.25\,\Delta T)$.
Overall, the data agree reasonably well with the predicted scaling.

To determine $\Psi^\dagger$, we must know both the isopycnal slope and the buoyancy gradient along the water–ice interface, the latter of which contains both temperature and salinity contributions (Eq.~\ref{eq:buoyancy}). Because $\Delta T$ is set by the ice-thickness variation $\Delta H_i$ (Eq.~\ref{eq:DeltaT}), two additional constraints are required to close the system and determine the under-ice salinity contrast $\Delta S$ and the isopycnal slope $s$. These constraints are obtained by requiring the tracer distributions to be in equilibrium.

First, to sustain a poleward thinning ice geometry, equatorial regions need to freeze and polar regions need to melt. The associated salinity flux into the ocean request a meridional salinity transport to balance.
Similarly, the equilibrium isopycnal slope $s$ must ensure that the downward buoyancy flux from diffusion balances the upward transport by baroclinic eddies. As shown by \citeA{Jansen-Kang-Kite-et-al-2023:energetic}, this upward buoyancy transport reflects a release of gravitational potential energy via a lowering of the system’s center of mass. When vertical buoyancy fluxes are balanced, the energy input from heat and salinity fluxes at the ice-water interface and from diffusion is exactly offset by the potential energy released through baroclinic eddy activity.

To analytically present the aforementioned balances, we write down the zonally integrated tracer equation and require the temporal tendency to vanish,
\begin{equation}
\label{eq:transport-eqn}
    0=\partial_t X=\mathcal{T}(X)\equiv \underbrace{(2\pi a\rho_0\cos\theta)\kappa_v \partial_z^2X}_{\rm diffusion}+\underbrace{\partial_z(\Psi^{\dagger} \partial_yX) -\partial_y(\Psi^{\dagger} \partial_zX)}_{\rm advection}.
  \end{equation}
   Here, $X$ can be either buoyancy $b$, temperature $T$ or salinity $S$. We multiply Eq.~\eqref{eq:transport-eqn} by an arbitrary test function $\phi$ and integrate by part to obtain the weak form. Here, we consider two test functions: 1) $\phi_1=a\theta$ ($\theta$ denotes latitude) for salinity transport and 2) $\phi_2=z-z_{\rm top}$ for buoyancy transport, to get the following necessary conditions for $b$ and $S$ profiles to be in equilibrium state,
\begin{eqnarray}
  \label{eq:salinity-flux-balance}
  &\underbrace{ -\int_0^{\pi / 2} \theta S_0 q \rho_0\left(2 \pi a^2\right) \cos \theta d \theta}_{\equiv \left.\Delta\mathcal{F}_{\rm S,v}\right|_{\rm top}}&=\underbrace{\left|\Psi^{\dagger}_0\right| \Delta S/\xi \cdot c }_{ \mathcal{F}_{\rm S,h}},\\
  \label{eq:vertical-transport-balance}
  &|\Psi^{\dagger}_0| \left(s+\rm{sign}(\Delta b)s_{\rm top}\right)&=\pi a\rho_0\kappa_v.
\end{eqnarray}
where $\xi\equiv  \rm{max}\{2sa/D,1\}$, $S_0$ is the ocean's mean salinity, and $q$ is the freezing/melting rate of the ice needed to balance the ice flow. The ice flow model is presented in SI Text S2. $c$, defined in SI Text S3, is a geometric factor that varies between $0.73$ and $2$.
Without delving into the derivations (see SI Text S3–S4), we briefly outline the physical processes embodied in Eqs.~\eqref{eq:salinity-flux-balance} and \eqref{eq:vertical-transport-balance}.

Eq.\eqref{eq:salinity-flux-balance} concerns the meridional transport of salinity.
Its right-hand-side $\mathcal{F}_{\rm S,h}$ represents the meridional salinity flux carried by baroclinic eddies. The surface term on the left, $\left.\Delta\mathcal{F}_{\mathrm{S},v}\right|_{\mathrm{top}}$ represents the resultant differential salinity flux between low and high latitudes,
\begin{equation}
\label{eq:salinity-flux-difference}
\left.\Delta\mathcal{F}_{\mathrm{S},v}\right|_{\mathrm{top}}\sim S_0 \rho_0 (q_{\rm eq}-q_{\rm pole}) (\pi a^2),
\end{equation}
where $q_{\mathrm{eq}}$ and $q_{\mathrm{pole}}$ are the freezing rates at the equator and pole respectively.
Equating $\left.\Delta\mathcal{F}_{\mathrm{S},v}\right|_{\mathrm{top}}$ and $\mathcal{F}_{\rm S,h}$, Eq.\eqref{eq:salinity-flux-balance} ensures that baroclinic eddies can carry the salt released by equatorial freezing toward the poles, to offset the freshening produced by polar melting.
Fig.\ref{fig:scaling-scatter}D shows that numerical simulations support this relationship. 

Eq.\eqref{eq:vertical-transport-balance} concerns the balance of vertical tracer transport. It ensures that downward buoyancy transport by diffusion ($\kappa_v b_z$) balances the upward buoyancy flux by eddies/residual circulation ($\Psi^\dagger_0b_y$, and $b_y$ can be replaced by $s b_z$), as well as the equivalent transport arising from buoyancy being consumed and replenished at different elevations through interaction with the ice ($|\Psi^\dagger_0|s_{\rm top}X_z$). Fig.\ref{fig:scaling-scatter}C shows that numerical simulations support this relationship. 

Jointly solving Eq.~\eqref{eq:Psi}, Eq.~\eqref{eq:salinity-flux-balance} and Eq.~\eqref{eq:vertical-transport-balance} gives the solution of $\Delta S$, $s$ and $\psi^\dagger_0$, which then can be used to evaluate the meridional heat transport (see SI Text S3)
\begin{equation}
  \label{eq:OHT}
  \mathcal{F} \equiv C_p \mathcal{F}_{ T,h}=C_p\left|\Psi^{\dagger}_0\right| \Delta T \xi^{-1} c .
\end{equation}
Shown in Fig.~\ref{fig:scaling}A-C are the $\Delta S$, $|\Psi^\dagger_0|$ and $\mathcal{F}$ solutions for a range of $S_0$ and $\kappa_v$, assuming Enceladus planetary parameters, as summarized in Table.\ref{tab:parameters}.

For a given mean ocean salinity $S_0$, increasing the vertical diffusivity $\kappa_v$ enhances the downward diffusion of surface temperature and salinity anomalies. This deepens isopycnals and strengthens the interior meridional density gradient, which in turn drives stronger circulation (Fig.\ref{fig:scaling}B) and enhances heat transport (Fig.\ref{fig:scaling}C). As a result, the equator-to-pole salinity contrast $\Delta S$ decreases (Fig.\ref{fig:scaling}A). When $\kappa_v$ becomes large enough that isopycnals reach the seafloor, vertical temperature and salinity contrasts are reduced (compare Fig.~\ref{fig:scaling}G2,3 with G1,4), suppressing meridional transport and altering the scaling relations in Eqs.\eqref{eq:Psi} and \eqref{eq:salinity-flux-balance}.

Fixing $\kappa_v$, the mean ocean salinity $S_0$ controls the direction and strength of circulation. In the low-salinity limit (Fig.\ref{fig:scaling}G1,2), salinity fluxes from ice-ocean exchange are weak, and thermal forcing dominates.
Since the thermal expansion coefficient $\alpha_T$ is negative in this regime, density increases poleward beneath the ice, driving sinking at the poles.
In the high-salinity limit (Fig.\ref{fig:scaling}G3,4), the anomalous thermal expansion is suppressed ($\alpha_T > 0$), and density increases equatorward, driving sinking at the equator.
At intermediate $S_0$, temperature- and salinity-induced buoyancy gradients partially cancel, resulting in weaker circulation, steeper isopycnals and reduced heat transport, consistent with \citeA{Kang-Mittal-Bire-et-al-2022:how}.

The two-dimensional parameter space spanned by $\kappa_v$ and $S_0$ thus can be divided into four circulation regimes: (1) low-salinity, low-diffusivity; (2) low-salinity, high-diffusivity; (3) high-salinity, high-diffusivity; and (4) high-salinity, low-diffusivity. These are illustrated in Fig.\ref{fig:scaling}G1–4, which show the characteristic balances and density structures. The transition between fresh and salty regimes (1–2 vs. 3–4) is marked by a reversal in circulation direction, indicated by a black dashed curve in Fig.\ref{fig:scaling}A,B. The transition between low- and high-diffusivity regimes (1,4 vs. 2,3), where isopycnals begin to outcrop at the seafloor, is marked by a gray solid curve in Fig.~\ref{fig:scaling}A,B.


It is evident from Fig.~\ref{fig:scaling}B,C that as salinity increases or decreases away from the reversal point, the ocean circulation $|\Psi^\dagger_0|$ and heat transport $\mathcal{F}$ both increase. However, the increase toward lower salinity is much stronger than toward higher salinity, which is somewhat counterintuitive because the rate at which $\alpha_T$ increases with salinity is nearly constant. One factor contributing to this asymmetry is the different geometric configurations in the salty-ocean and fresh-ocean scenarios. As shown in Fig.~\ref{fig:scaling}G, isopycnals are more widely spaced in the salty-ocean scenario because the water–ice interface tilts upward toward the poles, causing isopycnals to bend and thereby reducing the diffusive flux (Eq.~\ref{eq:vertical-transport-balance}), which in turn weakens $|\Psi^\dagger_0|$ and $\mathcal{F}$. In contrast, in the fresh-ocean scenario, isopycnals are compressed by topography, enhancing both the circulation and the heat transport. These results are consistent with \cite{Zhang-Kang-Marshall-2025:how}, except that only subcritical ($\xi=1$) cases are considered there.

Finally, we note that, in the low salinity regime where $\Psi^\dagger_0<0$, there exists a different set of solution with positive $\Psi^\dagger_0$, as shown in Fig.~\ref{fig:scaling}D-F (only unshaded regions are different from Fig.~\ref{fig:scaling}A-C). These solutions feature weak ocean circulation (small $|\Psi^\dagger_0|$), which is achieved by having salinity-induced density variations $\beta_S\Delta S$ almost exactly cancel out with the temperature-induced ones $\alpha_T\Delta T$. Similar bi-equilibrium states have been found to exist in idealized ocean circulation model for Earth ocean when both temperature and salinity forcings are present \cite{Stommel-1961:thermohaline}. However, in the context of icy moon ocean circulation, exact cancellation between $\beta_S\Delta S$ and $\alpha_T\Delta T$ is usually difficult to achieve, as pointed out by \cite{Kang-Mittal-Bire-et-al-2022:how}. This is because, unlike the under-ice temperature, which co-varies with the ice shell thickness $H_i$, the salinity flux profile is proportional to the second-order derivative of $H_i$ (SI Text S2). The different forcing profiles for temperature and salinity limit the degree of their cancellation. Therefore, we think the first set of solution shown in Fig.~\ref{fig:scaling}A-C is more relevant.

\section{3D numerical simulations for Enceladus ocean.}
\label{sec:numerical}
\begin{figure}[hpt!]
\centering \includegraphics[width=\textwidth, page=3]{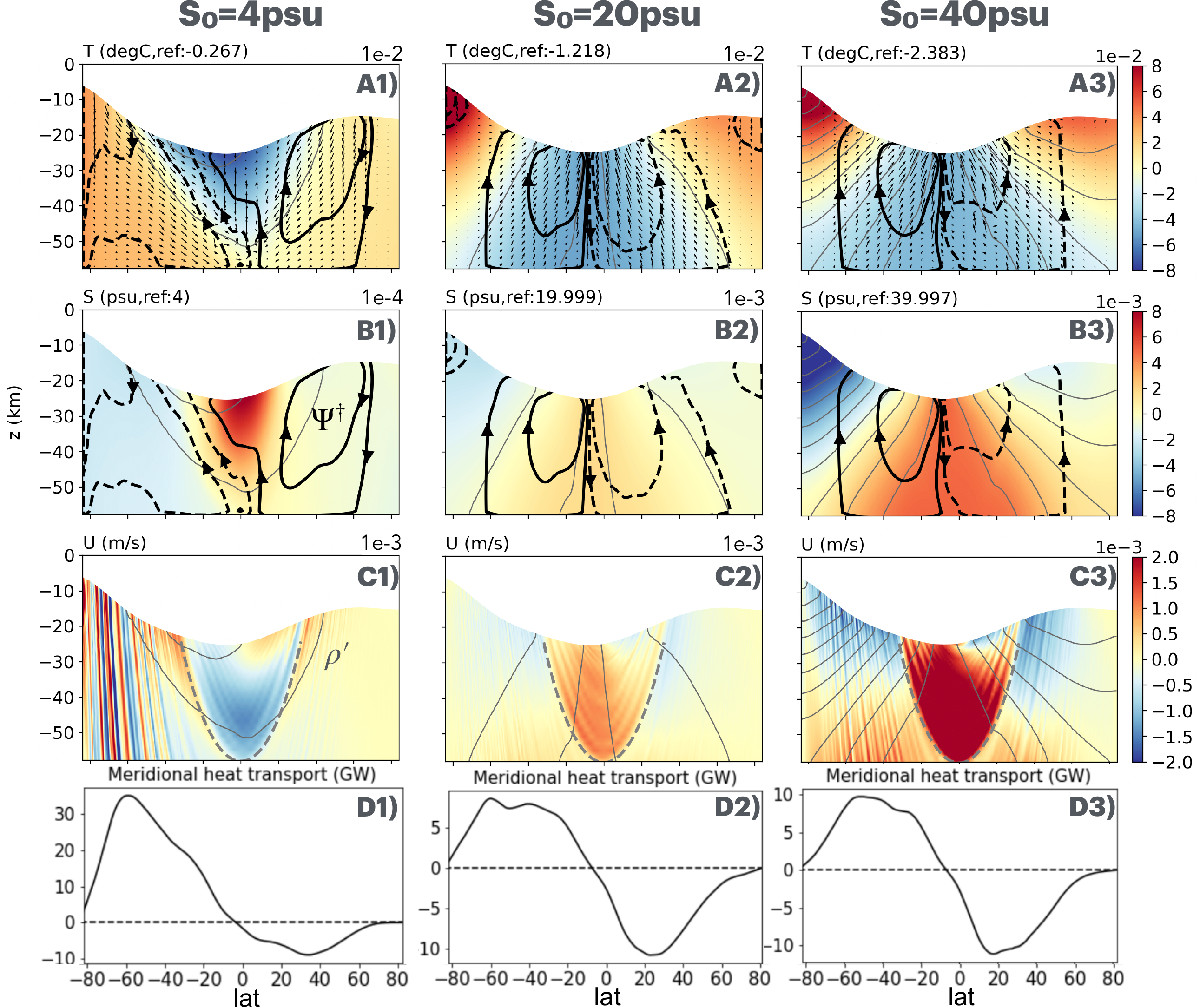}
\caption{Numerical solutions for the three high-diffusivity ($\kappa_v=10^{-2}$~m$^2$/s) simulations with different salinities. Shadings in panels (A-C) show the zonal-mean time-mean temperature $T$, salinity $S$ and zonal flow $U$ respectively. Thin gray contours in each panel present density. The spacing between two adjacent contours is set to $2\times 10^{-3}$~kg/m$^3$, and density increases with depth in all cases. Thick black contours with arrows in panels (A,B) show the diagnosed residual circulation streamfunction $\Psi^\dagger$, and solid/dashed $\Psi^\dagger$ contours denote clockwise/counter-clockwise circulation, respectively. The contour levels are $\pm 1.2\times 10^7,\ \pm 5\times 10^7,\ \pm 2\times 10^8$~kg/s. The arrows in panel (A) present the diagnosed eddy heat transport. Panels (D) show the time-mean meridional heat transport with positive values denote northward heat transport. 
}
\label{fig:3D-solution}
\end{figure}

To demonstrate the qualitative trends suggested by the analytical model, we conduct six sets of numerical simulations using MITgcm \cite{Marshall-Adcroft-Hill-et-al-1997:finite}, which cover two different vertical diffusivities $\kappa_v=10^{-3},\ 10^{-2}$~m$^2$/s and three different ocean salinities $S_0=4,\ 20,\ 40$~psu. We adopt the model setup in \citeA{Kang-Mittal-Bire-et-al-2022:how} except all experiments are three-dimensional instead of two-dimensional and are run under higher resolution (0.25$^\circ$) to capture the geostrophic turbulence generated through baroclinic instability. In this setup, the ocean temperature just beneath the ice is relaxed toward the local freezing point, and the freezing/melting rate of the ice is prescribed such that it counterbalances the tendency induced by ice flow and maintains the morphology of the ice shell unchanged (Fig.~\ref{fig:H-q}). Heat and momentum are exchanged between the ice and water at a rate of $\gamma_T=10^{-5}$~m/s and $\gamma_M=10^{-3}$~m/s, respectively. To represent the mixing by boundary layer turbulence as well as stabilize the water-ice interface, the vertical diffusivity at the boundary is enhanced by a factor of 40 and decays rapidly to interior value within the top five grid points. Since the freezing/melting rate is prescribed, it will not respond to the heat exchange between ice and ocean. By so doing, we enforce the equilibrium state of the ice shell and cut off the feedback loop between freezing/melting and ocean circulation, which usually causes the model to deviate far from what is realistic \cite{Kang-Bire-Marshall-2022:role}. The planetary radius, gravity, ice shell morphology, rotation rate are all set to Enceladus values. With a poleward thinning ice shell morphology as observed on Enceladus, water under the thick equatorial ice should be colder due to the freezing point suppression, and it should also be saltier assuming the freezing/melting rate of the ice can counterbalance the poleward ice flow. Further details of the model configuration can be found in the Materials \& Methods section in \citeA{Kang-Mittal-Bire-et-al-2022:how}, and the parameter choices are summarized in Table.\ref{tab:parameters}.

Conducting numerical simulations in this parameter regime is computationally expensive. To accelerate the convergence, we first conduct all simulations in a 2D setup representing the meridional plane for 50~kyr, and then use the final state to initialize a coarse resolution 3D simulation, which is run for another 500~yrs. During the integration, we stop the simulation from time to time and step to check how temperature and salinity evolves in the past few decades, and use the trend to step the temperature and salinity fields forward, to accelerate convergence. The final states of the coarse 3D simulations are used to initialize the presented simulations, and run for another 500~years with acceleration scheme. In the end, we make sure that the meridional convergence/divergence of heat and salinity flux matches up with the ocean-ice heat and salinity exchange. 

Shown in Fig.~\ref{fig:3D-solution} are the time-mean zonal-mean temperature $T$ (shading in panels A), salinity $S$ (shading in panels B) and density anomalies (contours in panels C). Across all experiments, temperature increases and salinity decreases from the equator to the poles, consistent with the changes of freezing point and the prescribed freezing/melting pattern (Fig.~\ref{fig:H-q}). While density always increases with depth, the meridional density gradient is opposite in the low salinity scenario ($S_0=4$~psu) and the high salinity scenarios ($S_0=20,\ 40$~psu), driving residual circulation in opposite directions. Residual circulation streamfunction $\Psi^\dagger$ (thick black contours in Fig.~\ref{fig:3D-solution}A,B) is diagnosed as
\begin{equation}
  \label{eq:residual-circulation-diag}
  \Psi^\dagger=\underbrace{\int_{\mathrm{bot}}^z (2\pi a\cos\phi) \rho \overline{v}~dz}_{\mathrm{Eulerian\ circulation\ by\ overturning\ cell}}+ \underbrace{(2\pi a\cos\phi) \rho \overline{w'T'}/\overline{T_y}}_{\mathrm{eddy-induced\ circulation}},
\end{equation}
where $\overline{(\cdot)}$ denotes zonal,time average, prime denotes deviation from the average, $\phi$ denotes latitude and $y$ denotes meridional distance. The residual circulation contains two components, one related to meridional overturning motions (Eulerian) and the other induced by the transport by baroclinic eddies.
The residual circulation always circulates dense fluid downward and buoyant fluid upward to lower the center of the mass of the fluid and release gravity potential energy.

Besides a reversed circulation, the opposite meridional density gradients in the low and high salinity scenarios also lead to opposite zonal flow patterns, shown by the shadings in Fig.~\ref{fig:3D-solution}C, consistent with thermal wind balance. It has been shown in many previous works that Enceladus ocean dynamics is strongly modulated by planetary rotation and thermal wind balance is well satisfied \cite{Bire-Kang-Ramadhan-et-al-2022:exploring, Kang-Mittal-Bire-et-al-2022:how}. Regardless of the direction of the circulation, heat is always converged toward the equator along the isopycnals (small black arrows in Fig.~\ref{fig:3D-solution}A), where water is generally colder. The vertically integrated heat transport is presented in Fig.~\ref{fig:3D-solution}D. 

\begin{figure}[hpt!]
\centering \includegraphics[width=\textwidth, page=4]{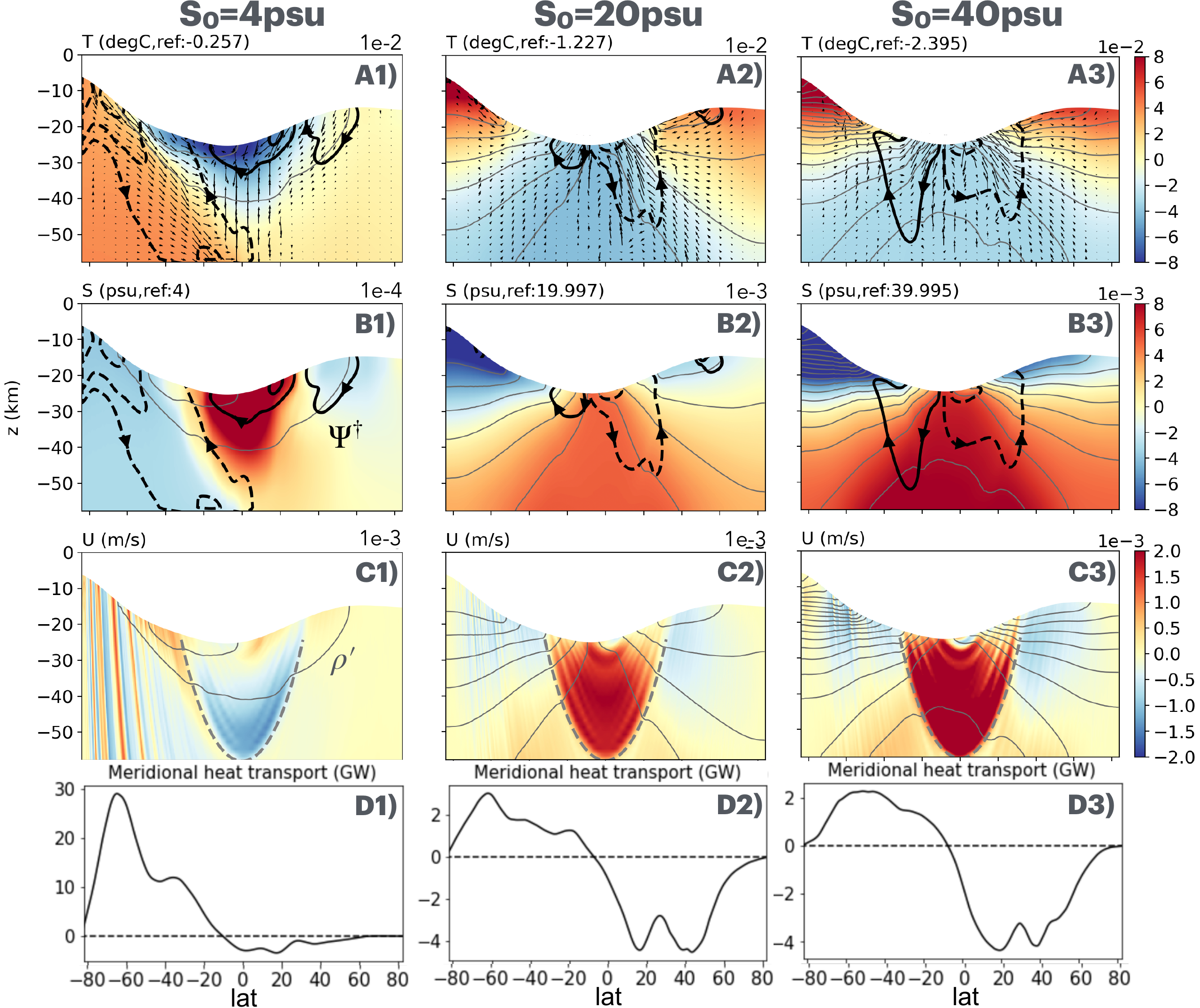}
\caption{Same as Fig.~\ref{fig:3D-solution}A-D, except for the lower-diffusivity ($\kappa_v=10^{-3}$~m$^2$/s) simulations.  }
\label{fig:3D-solution-lowdiff}
\end{figure}

Shown in Fig.~\ref{fig:3D-solution-lowdiff} are the solutions from the three lower-diffusivity simulations. Compared to the high diffusivity cases (Fig.~\ref{fig:3D-solution}), the penetration depth of the surface temperature (shading in panel A), salinity (shading in panel B) and density anomalies (contour in panel C) is less, resulting in weaker circulations (contour in panels A,B), weaker meridional heat transport (panel D) and weaker thermal-winds (shading in panel C). All these trends are qualitatively in line with the analytical prediction discussed in section~\ref{sec:scaling}.

To make qualitative comparison with scaling laws presented in section~\ref{sec:scaling}, we diagnose the equator-to-pole salinity contrast $\Delta S$, the global circulation strength $|\Psi^\dagger|$ and the equatorward ocean heat transport $\mathcal{F}$ from each simulation and overlay the results on Fig.~\ref{fig:scaling}. For $\Delta S$, we compute the difference between the maximum and minimum salinity in the northern and southern hemispheres, average between the two hemispheres. For the residual circulation strength, we compute the global mean $|\Psi^\dagger|$, and for ocean heat transport, we compute peak $|\mathcal{F}|$ values in each hemisphere and take average. As can be seen from Fig.~\ref{fig:scaling}, the numerical results match the theoretical prediction up to a factor of $3$. Features such as the enhancement of circulation and heat transport by strong diffusivity $\kappa_v$, the circulation reversal at low salinity, and the rapid strengthening of circulation and heat transport toward the low-salinity limit are all reproduced.

As a separate note, we did try to spin up the low-salinity, low-diffusivity case ($S_0=4$~psu, $\kappa_v=10^{-3}$~m$^2$/s) with a strong salinity gradient in the initial condition, in the hope to reproduce the other branch of solution (shown in Fig.~\ref{fig:scaling}D-F), which features sinking motions near the equator. However, after integrating the model for about 200 years, the circulation reverses, suggesting that the second solution may be hard to achieve in practice, if at all possible, due to the imperfect cancellation of temperature and salinity forcings. 

\section{Connections with observation and tidal modeling.}
\label{sec:obs-implications}
\subsection{Observed ice thickness variations may provide constraints on ocean tidal dissipation and ocean mean salinity.}
\label{sec:obs-oht}
In order to sustain the observed ice shell morphology on Enceladus, the equatorward heat convergence cannot be arbitrarily strong. Especially, if the converged heat flux is greater than the conductive heat loss rate at the equator, which is estimated to be no more than 8~GW 30$^\circ$S-30$^\circ$N \cite{Nimmo-Barr-Behounkova-et-al-2018:thermal}, the equatorial ice shell will necessarily melt. The melting combined with the poleward ice flow (SI Text S2) will make the equatorial ice shell get thinner over time instead of remaining stable. In fact, since ice is flowing poleward, the equatorial ice shell must be freezing to maintain its thickness. The latent heat release induced by freezing $\mathcal{H}_{\mathrm{latent}}$, the equatorward heat convergence in the ocean $\mathcal{F}$ and the conductive heat loss $\mathcal{H}_{\mathrm{cond}}$ should be in balance. As shown by the Fig.1d in \citeA{Kang-Mittal-Bire-et-al-2022:how}, if ice viscosity is set to $10^{14}$~Pa$\cdot$s at freezing point, $\mathcal{H}_{\mathrm{latent}}$ almost comparable to $\mathcal{H}_{\mathrm{cond}}$ at the equator, further lowering the upper bound for $\mathcal{F}$.

In Fig.~\ref{fig:scaling}C, three meridional heat transport contours are plotted, representing $\mathcal{F}=$1GW, 2GW, 4GW respectively. These values correspond to equatorward heat convergences of 2GW, 4GW, and 8GW as heat convergence accounting for the contribution from both hemispheres. The parameter regimes above these contours are likely to have too strong an OHT to be compatible with the observed ice shell morphology \cite{Hemingway-Mittal-2019:enceladuss,Park-Mastrodemos-Jacobson-et-al-2024:global, Schenk-McKinnon-2024:new}. This suggests that Enceladus ocean should have a vertical diffusivity $\kappa_v<10^{-3}$~m$^2$/s unless the ocean salinity is around 10~psu, which happens to coincide with \citeA{Kang-Mittal-Bire-et-al-2022:how}, who study the same problem using a two-dimensional model. While the conclusions here are aligned with \citeA{Kang-Mittal-Bire-et-al-2022:how}, it is worth noting that the ocean circulation depicted in this study arises primarily from baroclinic eddies, instead of boundary currents sustained by rough water-ice and water-rock interfaces \cite{Kang-Mittal-Bire-et-al-2022:how} --- a mechanism that is found to be less important under Enceladus parameter \cite{Zhang-Kang-Marshall-2024:ocean,Kang-2022:different}.
Also, it is worth noting that the scenarios with reversed circulation (sinking over the poles) tend to drive very strong OHT, which is unlikely to be compatible with the observed ice geometry on Enceladus. Similar results have been found by \citeA{Kang-Mittal-Bire-et-al-2022:how} and \citeA{Zeng-Jansen-2024:effect}.
The constraint on OHT also leads to a constraint on the circulation rate. Reading from Fig.~\ref{fig:scaling}B, the maximum feasible $\Psi^\dagger$ is around a few $10^7$~kg/s, which yields a circulation timescale of $>20$~kys.

\begin{figure}[hpt!]
\centering \includegraphics[width=\textwidth, page=7]{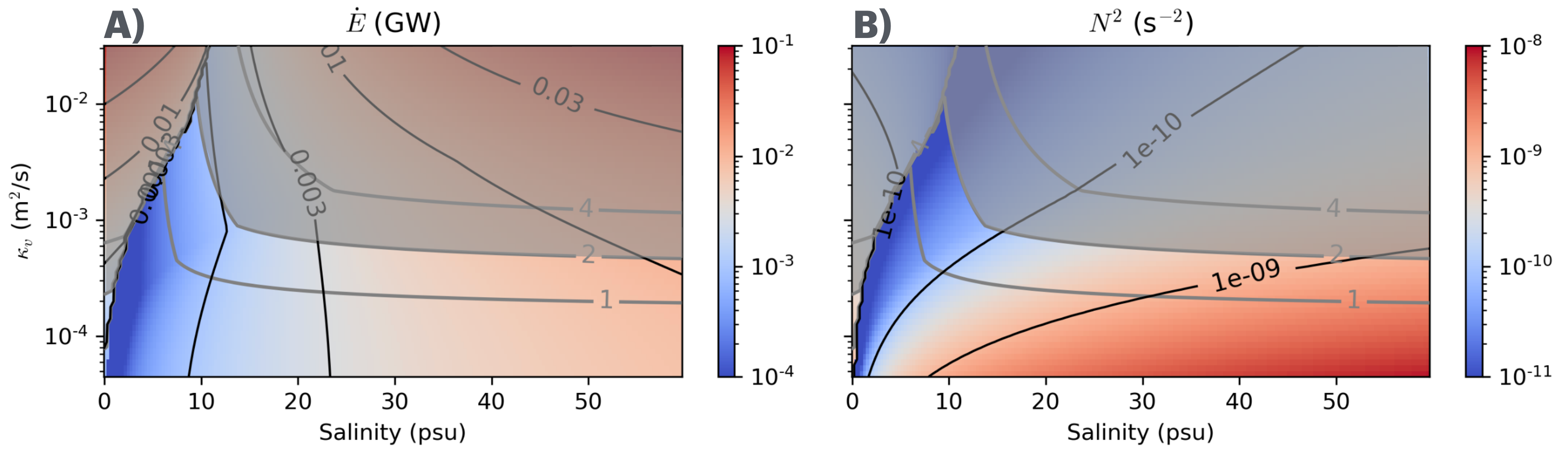}
\caption{Total ocean dissipation $\dot{E}$ and the stratification $N^2$ in the stratified upper ocean, predicted by the analytical model, for various ocean mean salinities $S_0$ and vertical diffusivity $\kappa_v$. Black contours present isolines of $\dot{E}$ and $N^2$, and the gray contours present the OHT $\mathcal{F}$ shown in Fig.~\ref{fig:scaling}C. High OHT (gray shading) can destroy the observed poleward thinning ice thickness profile on Enceladus over time. }
\label{fig:dissipation-N2}
\end{figure}

Vertical diffusivity $\kappa_v$ is contributed by molecular motions and mixing induced by tidal wave breaking \cite{Osborn-1980:estimates} and convective plume breaking \cite{Lecoanet-Quataert-2013:internal}. Therefore, upper bounds on $\kappa_v$ can be converted into constraints on the tidal dissipation rate, neglecting the mixing due to other processes (molecular diffusion and convection). 
The tidal dissipation process has been studied using Earth-based observations, numerical simulations and theoretical analysis \cite{Peltier-Caulfield-2003:mixing}.

Generally speaking, the dissipated kinetic energy is partially converted into heat and partially into gravity potential energy by mixing dense fluid upward, and the proportion that goes into gravity potential energy is roughly $\Gamma=20\%$ if the wave breaking occurs through Kelvin-Helmholtz instability, leading to the famous Osborn relationship \cite{Osborn-1980:estimates,Wunsch-Ferrari-2004:vertical},
\begin{equation}
  \label{eq:Osborn}
  \kappa_v=\Gamma \frac{\epsilon}{\partial_z b}=\frac{\Gamma\epsilon 2a (s+\rm{sign}(\Delta b) s_{\rm top})}{\Delta b}.
\end{equation}
Here, the stratification $\partial_z b$ is rewritten as the under-ice buoyancy contrast $\Delta b$ (defined by Eq.~(\ref{eq:buoyancy}) divided by the penetration depth $2a(s+\rm{sign}(\Delta b) s_{\rm top})$.
Substituting Eq.\ref{eq:Osborn} into Eq.~(\ref{eq:OHT}) and Eq.~(\ref{eq:vertical-transport-balance}), we get
\begin{equation}
  \label{eq:heat-trans-Osborn}
  \mathcal{F}=\frac{2\pi a^2C_p\rho_0\Gamma\epsilon c}{\alpha_T g}\frac{1}{1+(\beta_S/\alpha_T)(\Delta S/\Delta T)}\xi^{-1},
\end{equation}
where $\xi \equiv \max \{2 \mathrm{sa} / \mathrm{D}, 1\}$. We then recognize that the residual streamfunction $\Psi^\dagger_0$ should act on both temperature and salinity gradient indifferently, which allows us to replace $\Delta S/\Delta T$ with $\mathcal{F}_{S,h}/\mathcal{F}_{T,h}=    \left.\Delta\mathcal{F}_{\rm S,v}\right|_{\rm top}/\mathcal{F}_{T,h}$,
\begin{equation}
  \label{eq:heat-trans-Osborn-cancel-dS}
  \mathcal{F}=\frac{2\pi a^2 C_p\rho_0\Gamma \epsilon c}{\alpha_T g}\xi^{-1}-(\beta_S/\alpha_T)C_p\left.\Delta\mathcal{F}_{\rm S,v}\right|_{\rm top},
\end{equation}
where $\left.\Delta\mathcal{F}_{\rm S,v}\right|_{\rm top}$ is defined in Eq.~(\ref{eq:salinity-flux-balance}) to represent the equator-to-pole difference of salinity flux needed to balance the ice flow.

Figure~\ref{fig:dissipation-N2}A shows the total ocean dissipation required, $\dot{E} = \rho_0 \epsilon (4\pi a^2 D)$, to sustain the assumed vertical diffusivity $\kappa_v$ across the $\kappa_v$-$S_0$ parameter space. Gray curves indicate isolines of equatorward ocean heat transport $\mathcal{F}$ at 1, 2, and 4~GW. The region where $\mathcal{F} > 2$~GW is shaded, as such strong heat transport may be incompatible with the observed poleward-thinning ice shell on Enceladus. For combinations of $\kappa_v$ and $S_0$ that yield $\mathcal{F} < 2$GW, the corresponding ocean dissipation $\dot{E}$ remains below 0.01GW, which is negligible compared to Enceladus’s overall heat budget. Even smaller $\dot{E}$ is obtained at low salinity. 

The extremely low $\dot{E}$ can be understood through the concept of heat engine. Let us first consider a simple scenario where salinity flux $\left.\Delta\mathcal{F}_{\rm S,v}\right|_{\rm top}$ is negligible\footnote{Salinity flux tends to be less important when ice viscosity is high or if a large satellite is considered \cite{Kang-Jansen-2022:icy}.} and diffusivity is sufficiently low for isopycnal to avoid touching the seafloor ($\xi=1$). In this case, Eq.\ref{eq:heat-trans-Osborn-cancel-dS} reduces to $\frac{2\pi a^2 C_p\rho_0\Gamma \epsilon c}{\alpha_T g}$. We may divide $\mathcal{F}$ over the total dissipation in the ocean $\dot{E}=\rho_0\epsilon (4\pi a^2D)$ (assuming dissipation is uniform) to estimate the OHT induced per 1W of dissipation.
\begin{equation}
  \label{eq:heat-ratio-Osborn-xi1-noSflux}
  \frac{2\mathcal{F}}{\dot{E}}=\frac{2C_p\Gamma c }{D\alpha_T g}\sim O(10^3).
\end{equation}
With $\Gamma\sim 0.2$ \cite{Ivey-Imberger-1991:nature}, the dissipation-to-OHT yield ratio is in the order of $10^3$ --- every one Watt of heat dissipated in the ocean has the potential to induce a thousand Watts of heat redistribution. This high heat transport efficiency arises from the same principle that allows heat pump HVAC systems to transfer more heat than the amount of work they consume. For Enceladus, to keep the OHT below a few GW, as indicated by the observed ice thickness profile, the total ocean dissipation should be less than a few MW following Eq.\ref{eq:heat-ratio-Osborn-xi1-noSflux}. 

Next, we consider the scenarios with isopycnals touching the seafloor. With $\xi>1$, the expression of OHT $\mathcal{F}$ will need to be modified by a correction factor $\xi^{-1}$ that is always less than unity. The reduction of OHT arises from the limitation of the domain depth. 
Thanks to this correction, the upper bound of ocean dissipation rate $\epsilon$ will be raised for the intermediate salinity scenarios, where $\xi>1$ is achieved (Fig.~\ref{fig:scaling}G2,G3). 

Finally, as can be seen from Eq.~(\ref{eq:heat-trans-Osborn-cancel-dS}), salinity flux $\left.\Delta\mathcal{F}_{\rm S,v}\right|_{\rm top}$ (defined in Eq.~\ref{eq:salinity-flux-balance}) also decreases the OHT and raises the upper bound for ocean dissipation under Enceladus setup. This can be understood from an energetic point of view. As demonstrated in \citeA{Jansen-Kang-Kite-et-al-2023:energetic}, since the equatorial freezing occurs at a lower elevation than the polar melting, the net energy input to ocean energetics associated with freezing/melting is negative. Diffusion first needs to overcome the energy consumption by $    \left.\Delta\mathcal{F}_{\rm S,v}\right|_{\rm top}$ before it can drive ocean dynamics and OHT. The forgiven amount of ocean dissipation $\dot{E}_S=\frac{2\beta_S gD \left.\Delta\mathcal{F}_{\rm S,v}\right|_{\rm top}}{\rho_0\Gamma }\xi$ is merely 2~kW, assuming $\xi=1$ and an ice viscosity of $\eta_m=10^{14}$~Pa$\cdot$s at freezing point, which is again negligible compared to the total heat production on Enceladus. 

According to Eq.~(\ref{eq:heat-trans-Osborn-cancel-dS}), if ocean dissipation $\dot{E}$ is vanishingly small, the OHT can be negative in presence of salinity flux. This corresponds to a scenario, where salinity contrast between the equator and poles $\Delta S$ secularly builds up, due to the absence of ocean circulation. Accounting for molecular diffusion may allow the system to equilibrate. 
The analysis presented here assumes zero heat flux from the silicate core, however, core dissipation has been proposed in many previous works \cite{Roberts-2015:fluffy, Choblet-Tobie-Sotin-et-al-2017:powering} to be a key heat source to keep the ocean from freezing, given the insufficient dissipation estimated for the ice shell \cite{Beuthe-2019:enceladuss, Soucek-Behounkova-Cadek-et-al-2019:tidal, Robuchon-Choblet-Tobie-et-al-2010:coupling, Shoji-Hussmann-Kurita-et-al-2013:ice, Behounkova-Tobie-Choblet-et-al-2013:impact, McCarthy-Cooper-2016:tidal, Beuthe-2019:enceladuss, Soucek-Behounkova-Cadek-et-al-2019:tidal, Gevorgyan-Boue-Ragazzo-et-al-2020:andrade}. Recent studies by \citeA{Liao-Nimmo-Neufeld-2020:heat} and \citeA{Rovira-Navarro-Katz-Liao-et-al-2022:tides} attempt to build self-consistent models for porous-viscoelastic media, and they have concluded that significant heat generation in the silicate core is only feasible if the core's rigidity is low.
When the heat flux from the silicate core is nonzero, this heat is likely to all be deflected toward the equatorial ice shell by the stratified layer under the polar ice shell (Fig.~\ref{fig:3D-solution}C3, Fig.~\ref{fig:3D-solution-lowdiff}C3), regardless of their distribution at the seafloor \cite{Kang-2023:modulation}. This indicates that, equatorial regions will receive more heat in presence of core heating, which will further narrow down the parameter regime that is consistent with the observed ice thickness profile, as depicted in Fig.~\ref{fig:scaling}. More research is needed to account for the influence of bottom heat flux.

\subsection{The feedback between tides and baroclinic eddies.}
\label{sec:obs-tide}

Tidal processes and baroclinic eddies are dynamically coupled. On one hand, the tidal dissipation rate $\dot{E}$ depends on the ocean’s stratification $N^2$, which influences the properties of internal gravity waves excited by tides \cite{Tyler-2020:heating, Rovira-Navarro-Rieutord-Gerkema-et-al-2019:do, Hay-Matsuyama-2019:nonlinear, Rekier-Trinh-Triana-et-al-2019:internal, Idini-Nimmo-2024:resonant}. On the other hand, the stratification itself is shaped by a balance between vertical mixing (parameterized by $\kappa_v$) and transport by baroclinic eddies. Our analytical model captures this latter half of the feedback loop, offering computationally efficient predictions for $N^2$ given a prescribed $\kappa_v$.

Figure~\ref{fig:dissipation-N2}B shows the predicted equilibrium stratification, computed as $N^2 = |\Delta b|/(2sa + \mathrm{sign}(\Delta b)\Delta H)$ (with $\Delta b$ defined in Eq.\eqref{eq:buoyancy}), as a function of $\kappa_v$ and mean ocean salinity $S_0$. Two trends emerge: (1) as $\kappa_v$ and $\dot{E}$ (Fig.\ref{fig:dissipation-N2}A) decreases, the circulation would be able to push isopycnals close to the water-ice interface, enhancing the stratification; (2) intermediate salinities tend to yield weak stratification, as temperature- and salinity-driven buoyancy contributions partially cancel beneath the ice. The Brunt-Vaisala frequency $N$ may be comparable or even greater than the Coriolis frequency $f$ in the limit of weak tidal dissipation, allowing buoyancy-dominant internal gravity waves to form in the ocean \cite{Idini-Nimmo-2024:resonant}. 

In the coupled tide-circulation system, equilibrium states can be identified by finding the intersections of the $N^2(\dot{E})$ relation from our model and the $\dot{E}(N^2)$ relation from tidal theory. If the tidal response present resonant features \cite{Idini-Nimmo-2024:resonant}, multiple equilibria may exist, some stable, some unstable. The feedback between circulation and tidal forcing could then enable the system to remain near resonance.

\section{Conclusion and discussion.}
\label{sec:discussion}

Enceladus’s ocean is driven by both thermal and salinity forcing at the ice-water interface. The poleward-thinning ice creates pressure gradient at the water-ice interface and thereby a temperature gradient due to the dependence of water's freezing point on pressure. Maintaining this thickness gradient also requires net freezing near the equator and melting near the poles to balance poleward ice flow, driving a meridional salinity gradient. Together, these buoyancy forcings power ocean circulation and heat transport, energized by vertical mixing within the water column \cite{Jansen-Kang-Kite-et-al-2023:energetic}.
The vertical diffusivity $\kappa_v$ can greatly exceed the molecular value $\kappa_m$ accounting for the contributions from internal-tide breaking \cite{Osborn-1980:estimates} and convective-plume turbulence \cite{Lecoanet-Quataert-2013:internal}. On Earth, for example, $\kappa_v\sim 10^{-4},\mathrm{m^2,s^{-1}}$ -- two to three orders of magnitude larger than $\kappa_m$ -- largely due to tidal mixing \cite{Munk-Wunsch-1998:abyssal}.

In this work, we concern the ocean circulation and heat transport on Enceladus driven by the aforementioned thermal and saline forcings from the ice. We assume that the water temperature beneath Enceladus’s ice shell is near the freezing point, and that the freezing/melting of the ice (which sets the salinity flux into the ocean) exactly balances the ice flow, maintaining the poleward-thinning ice geometry (see Fig.~\ref{fig:H-q}a).
For this system, we develop analytical scaling laws for baroclinic eddies (Eq.\eqref{eq:Psi}), and combine it with the ocean’s heat and salinity budgets (Eqs.\eqref{eq:salinity-flux-balance}–\eqref{eq:vertical-transport-balance}) to predict the meridional heat transport $\mathcal{F}$, the overturning strength $\Psi^\dagger_0$, and the equator-to-pole salinity contrast $\Delta S$ and the ocean stratification $N^2$ for different ocean diffusivities $\kappa_v$ and ocean mean salinities $S_0$. These predictions are benchmarked against numerical simulations (Fig.~\ref{fig:scaling}).

We find that increasing $\kappa_v$ strengthens both $\Psi^\dagger_0$ and $\mathcal{F}$ while reducing $\Delta S$, because enhanced mixing erodes salinity gradients, consistent with \citeA{Kang-2022:different, Zhang-Kang-Marshall-2024:ocean}. Furthermore, changing ocean mean salinity can reverse the direction of the ocean circulation significantly alter the strength of ocean heat transport. Compared to salty ocean, fresh ocean tends to induce a much stronger $\mathcal{F}$ in the opposite direction, and intermediate ocean salinity tends to minimize heat transport, in line with \citeA{Kang-Mittal-Bire-et-al-2022:how}. 

Our scaling framework builds a connection between the more observable ice-shell geometry and the less constrained properties such as oceanic tidal dissipation, ocean stratification and mean salinity. This connection would enable tighter constraints on ocean parameters.
We proposed two possible scenarios to apply our theory (Fig.~\ref{fig:dissipation-N2}):
\begin{itemize}
\item Suppose the equator-to-pole ice-shell thickness difference is measured, it can be used to constrain on the meridional heat flux $\mathcal{F}$ assuming the ice heat budget is in balance \cite{Kang-Mittal-Bire-et-al-2022:how}. Using our model, the constraints on $\mathcal{F}$ can be turned into constraints on the parameters (such as the mean salinity $S_0$ and tidal dissipation rate $\dot{E}$) that controls $\mathcal{F}$ (see section~\ref{sec:obs-oht}).
\item Tides and baroclinic eddies form a coupled system. Tidal mixing (an important contributor to the vertical diffusivity $\kappa_v$) steepens isopycnals and energizes baroclinic eddies and ocean circulation. In turn, baroclinic eddies drive heat transport, shaping the ocean’s temperature and salinity structure and thereby influencing the tidal dissipation rate \cite{Tyler-2020:heating, Rovira-Navarro-Rieutord-Gerkema-et-al-2019:do, Hay-Matsuyama-2019:nonlinear, Rekier-Trinh-Triana-et-al-2019:internal,Idini-Nimmo-2024:resonant}. Our analytical model predicts the ocean stratification $N^2$ for a given $\dot{E}$, and can thus be coupled with tidal models to constrain possible equilibrium states (see section~\ref{sec:obs-tide}).
\end{itemize}


\begin{table}[hptb!]
  \centering
  \begin{tabular}{lll}
    \hline
    Symbol & Name & Definition/Value\\
    \hline
    \multicolumn{3}{c}{Physical constants}\\
    \hline
    $L_f$ & fusion energy of ice & 334000~J/kg\\
    $C_p$ & heat capacity of water & 4000~J/kg/K\\
    $\rho_i$ & density of ice & 917~kg/m$^3$ \\
    $\rho_0$ & density of seawater & 1000~kg/m$^3$ \\
    $f_b$ & sensitivity of water freezing point to pressure & $-7.6\times 10^{-8}$\,K/Pa \\
    $a$ & Enceladus radius & 252~km\\
        \hline
    \multicolumn{3}{c}{Variables} \\
        \hline
    $u$, $v$, $w$ & three-component velocity & prognostic \\
    $T$ & temperature & prognostic \\
    $S$ & salinity & prognostic \\
    $\mathcal{F}_{X,h}$ & horizontal transport of tracer $X$ by eddies
        & Eq.~\ref{eq:horizontal-vertical-fluxes}\\
    $\mathcal{F}_{X,v}$ & vertical transport of tracer $X$ by eddies
        & Eq.~\ref{eq:horizontal-vertical-fluxes}\\
    $\mathcal{F}$ & meridional heat transport & Eq.~\ref{eq:OHT} \\
        \hline
    \multicolumn{3}{c}{Analytical model} \\
    \hline
    $\Delta T$ & equator-to-pole under-ice temperature difference & Eq.~\ref{eq:DeltaT}\\
    $g_0$ & gravity constant & 0.113\,m/s$^2$ \\
    $\Delta H_i$ & equator-to-pole ice thickness difference & 15~km\\
    $\left.\Delta\mathcal{F}_{S,v}\right|_{\rm top}$ & equator-to-pole under-ice salinity
        flux difference
        & Eq.~\ref{eq:salinity-flux}, Eq.~\ref{eq:salinity-flux-difference} \\
    $q_\mathrm{eq} - q_\mathrm{pole}$ & equator-to-pole under-ice freezing rate difference
        & 0.6~km/Myr \\
    $\eta_m$ & ice viscosity at freezing point & 2$\times$10$^{14}$~Pa$\cdot$s (used to compute $q$)\\
    $S_0$ & mean ocean salinity & 2, 20, 40 psu \\
    $\Delta S$ & under-ice equator-to-pole salinity difference & Eq.~\ref{eq:salinity-flux-balance}\\
    $\Psi_0^\dagger$ & eddy-driven overturning circulation & Eq.~\ref{eq:Psi}\\
    $s$ & isopycnal slope & solved from Equations~\ref{eq:Psi}, \ref{eq:vertical-transport-balance} \\
    $\kappa_v$ & vertical thermal and salinity diffusivity & 0.001,0.01~m$^2$/s \\
    $\beta$ & meridional gradient of planetary vorticity
    & $4.2\times 10^{-10}\,\mathrm{m}^{-1}\,\mathrm{s}^{-1}$\\
    $s_\mathrm{top}$ & slope of the water-ice interface & $(\rho_i/\rho_o)(\Delta H_i / R) \approx 0.055$ \\
    $\Delta b$ & equator-to-pole under-ice buoyancy difference & Eq.~\ref{eq:buoyancy} \\
    $\alpha_T,\beta_S$ & thermal expansivity, saline contractivity & Gibbs Seawater Toolbox (ref 2)\\
    $k$ & diffusion constant in the mixing-length theory & 0.25 \\
    $\xi$ & factor for the effect of isopycnal-seafloor contact & $\max\{2sa/D, 1\}$\\
    $D$ & global mean ocean depth& 40~km (ref 1) \\
    $c$ & geometric factor between 0.73 and 2 & Supplementary (Eq.~18)\\
    \hline
    \multicolumn{3}{c}{Parameters in the numerical simulations}\\
    \hline
    $g$ & gravity in the ocean & Eq.8 in ref 4\\
    $\Omega$ & rotation rate & 5.307$\times$10$^{-5}$~s$^{-1}$\\
    $H_i$ & ice thickness & Fig.\ref{fig:H-q}, zonally average map by ref 1  \\
    $D+H_i$ & total depth of ocean and ice layer & 60~km \\
    $q$ & prescribed freezing rate & Eq.9 in SI, Fig.\ref{fig:H-q} \\
    $T_f$ & freezing point at water-ice interface & $T_f=c_0+b_0P+a_0S$\\
    $a_0,b_0,c_0$ & freezing point coefficients & $-0.058$K/psu, $-7.6$e-8K/Pa $0.09^\circ$C\\
    $\rho(T,S,P)$ & equation of state & MDJWF scheme (ref 3)\\
    $\nu_h,\ \nu_v$ & horizontal/vertical viscosity & 0.001,0.01~m$^2$/s\\
    $\tilde{\nu}_h,\ \tilde{\nu}_v$ & bi-harmonic hyperviscosity & 0~m$^4$/s\\ 
    $\kappa_h$ & horizontal diffusivity & $\kappa_v$\\
    $\gamma_T,\ \gamma_S,\ \gamma_M$ & water-ice exchange coeff. for T, S \& motion & 10$^{-5}$, 10$^{-5}$, 10$^{-3}$~m/s\\ 
    \hline
     \end{tabular}
  \caption{Parameters and variables used in this study. Ref 1: \cite{Hemingway-Mittal-2019:enceladuss}; ref 2: \cite{McDougall-Barker-2011:getting}; ref 3: \cite{McDougall-Jackett-Wright-et-al-2003:accurate}; ref 4: \citeA{Kang-Mittal-Bire-et-al-2022:how}}
  \label{tab:parameters}
  
\end{table}

\section*{Open Research Section}
The model setup used in this analysis is available at Zenodo, DOI: 10.5281/zenodo.19076565. 

\acknowledgments
This work was supported by the NASA ICAR award 80NSSC26K0263 and by the Research Committee Fund provided by MIT.

%

\bibliography{export}

@String{ARFM		=	"Ann. Rev. Fluid Mech."}

@String{EPSL		=	"Earth Planet. Sci. Lett."}

@String{GRL		=	"Geophys. Res. Lett."}

@String{JAS		=	"J. Atmos. Sci."}

@String{JGR		=	"J. Geophys. Res."}

@String{NATURE		=	"Nature"}

@String{JAMES           =       "J. Adv. Model. Earth Syst."}

@article{Ashkenazy-Sayag-Tziperman-2018:dynamics,
  author =	 {Ashkenazy, Yosef and Sayag, Roiy and Tziperman, Eli},
  title =	 {{Dynamics of the global meridional ice flow of
                  Europa{\textquoteright}s icy shell}},
  journal =	 {Nature Astronomy},
  year =	 2018,
  volume =	 2,
  number =	 1,
  pages =	 {43--49},
  month =	 jan
}

@article{Ashkenazy-Tziperman-2021:dynamic,
  title =	 {Dynamic Europa ocean shows transient Taylor columns
                  and convection driven by ice melting and salinity},
  author =	 {Ashkenazy, Yosef and Tziperman, Eli},
  journal =	 {Nature communications},
  volume =	 12,
  number =	 1,
  pages =	 {1--12},
  year =	 2021,
  publisher =	 {Nature Publishing Group}
}

@article{Behounkova-Tobie-Choblet-et-al-2013:impact,
  title =	 {Impact of tidal heating on the onset of convection
                  in Enceladus’s ice shell},
  author =	 {B{\v{e}}hounkov{\'a}, Marie and Tobie, Gabriel and
                  Choblet, Ga{\"e}l and {\v{C}}adek, Ond{\v{r}}ej},
  journal =	 {Icarus},
  volume =	 226,
  number =	 1,
  pages =	 {898--904},
  year =	 2013,
  publisher =	 {Elsevier}
}

@article{Beuthe-2019:enceladuss,
  title =	 {Enceladus's crust as a non-uniform thin shell: II
                  tidal dissipation},
  journal =	 {Icarus},
  volume =	 {332},
  pages =	 {66 - 91},
  year =	 {2019},
  issn =	 {0019-1035},
  doi =		 {10.1016/j.icarus.2019.05.035},
  author =	 {Mikael Beuthe}
}

@article{Beuthe-Rivoldini-Trinh-2016:enceladuss,
  author =	 {Beuthe, Mikael and Rivoldini, Attilio and Trinh,
                  Antony},
  title =	 {{Enceladus's and Dione's floating ice shells
                  supported by minimum stress isostasy}},
  journal =	 {GRL},
  year =	 2016,
  volume =	 43,
  number =	 19,
  pages =	 {10,088--10,096},
  month =	 oct
}

@article{Bire-Kang-Ramadhan-et-al-2022:exploring,
  title =	 {Exploring ocean circulation on icy moons heated from
                  below},
  author =	 {Bire, Suyash and Kang, Wanying and Ramadhan, Ali and
                  Campin, Jean-Michel and Marshall, John},
  journal =	 {JGR: Planets},
  pages =	 {e2021JE007025},
  year =	 2022,
  volume =	 127,
  month =	 mar,
  publisher =	 {Wiley Online Library}
}

@article{Carr-Belton-Chapman-et-al-1998:evidence,
  title =	 {Evidence for a subsurface ocean on Europa},
  author =	 {Carr, Michael H and Belton, Michael JS and Chapman,
                  Clark R and Davies, Merton E and Geissler, Paul and
                  Greenberg, Richard and McEwen, Alfred S and Tufts,
                  Bruce R and Greeley, Ronald and Sullivan, Robert and
                  others},
  journal =	 {Nature},
  volume =	 391,
  number =	 6665,
  pages =	 {363--365},
  year =	 1998,
  publisher =	 {Nature Publishing Group}
}

@article{Choblet-Tobie-Sotin-et-al-2017:powering,
  author =	 {Choblet, Ga{\"e}l and Tobie, Gabriel and Sotin,
                  Christophe and B{\v e}hounkov{\'a}, Marie and {\v
                  C}adek, Ond{\v{r}}ej and Postberg, Frank and Sou{\v
                  c}ek, Ond{\v{r}}ej},
  title =	 {{Powering prolonged hydrothermal activity inside
                  Enceladus}},
  journal =	 {Nature Astronomy},
  year =	 2017,
  volume =	 1,
  number =	 12,
  pages =	 {841--847},
  month =	 dec
}

@article{Chyba-2000:energy,
  title =	 {Energy for microbial life on Europa},
  author =	 {Chyba, Christopher F},
  journal =	 {Nature},
  volume =	 403,
  number =	 6768,
  pages =	 {381--382},
  year =	 2000,
  publisher =	 {Nature Publishing Group}
}

@article{Cockell-Simons-Castillo-Rogez-et-al-2023:sustained,
  title =	 {Sustained and comparative habitability beyond Earth},
  author =	 {Cockell, Charles S and Simons, Mark and
                  Castillo-Rogez, Julie and Higgins, Peter M and
                  Kaltenegger, Lisa and Keane, James T and Leonard,
                  Erin J and Mitchell, Karl L and Park, Ryan S and
                  Perl, Scott M and others},
  journal =	 {Nature Astronomy},
  pages =	 {1--9},
  year =	 2023,
  publisher =	 {Nature Publishing Group UK London}
}

@article{Gevorgyan-Boue-Ragazzo-et-al-2020:andrade,
  title =	 {Andrade rheology in time-domain. Application to
                  Enceladus' dissipation of energy due to forced
                  libration},
  author =	 {Gevorgyan, Yeva and Bou{\'e}, Gwena{\"e}l and
                  Ragazzo, Clodoaldo and Ruiz, Lucas S and Correia,
                  Alexandre CM},
  journal =	 {Icarus},
  volume =	 343,
  pages =	 113610,
  year =	 2020,
  publisher =	 {Elsevier}
}

@article{Glein-Waite-2020:carbonate,
  author =	 {Glein, Christopher R and Waite, J Hunter},
  title =	 {{The Carbonate Geochemistry of Enceladus' Ocean}},
  journal =	 {GRL},
  year =	 2020,
  volume =	 47,
  number =	 3,
  pages =	 591,
  month =	 feb
}

@article{Hand-Chyba-2007:empirical,
  author =	 {Hand, K and Chyba, C},
  title =	 {{Empirical constraints on the salinity of the
                  europan ocean and implications for a thin ice
                  shell}},
  journal =	 {Icarus},
  year =	 2007,
  volume =	 189,
  number =	 2,
  pages =	 {424--438},
  month =	 aug
}

@article{Hay-Matsuyama-2019:nonlinear,
  author =	 {Hay, Hamish C F C and Matsuyama, Isamu},
  title =	 {{Nonlinear tidal dissipation in the subsurface
                  oceans of Enceladus and other icy satellites}},
  journal =	 {Icarus},
  year =	 2019,
  volume =	 319,
  pages =	 {68--85},
  month =	 feb
}

@article{Held-Larichev-1996:scaling,
  title =	 {A scaling theory for horizontally homogeneous,
                  baroclinically unstable flow on a beta plane},
  author =	 {Held, Isaac M and Larichev, Vitaly D},
  journal =	 {JAS},
  volume =	 53,
  number =	 7,
  pages =	 {946--952},
  year =	 1996
}

@incollection{Hemingway-Iess-Tadjeddine-et-al-2018:interior,
  author =	 {Hemingway, D and Iess, L and Tadjeddine, R and
                  Tobie, G},
  title =	 {{The Interior of Enceladus}},
  booktitle =	 {Enceladus \& the Icy Moons of Saturn},
  year =	 2018,
  publisher =	 {The University of Arizona Press}
}

@article{Hemingway-Mittal-2019:enceladuss,
  author =	 {Hemingway, Douglas J and Mittal, Tushar},
  title =	 {{Enceladus's ice shell structure as a window on
                  internal heat production}},
  journal =	 {Icarus},
  year =	 2019,
  volume =	 332,
  pages =	 {111--131},
  month =	 nov
}

@article{Idini-Nimmo-2024:resonant,
  title =	 {Resonant stratification in titan’s global ocean},
  author =	 {Idini, Benjamin and Nimmo, Francis},
  journal =	 {The Planetary Science Journal},
  volume =	 5,
  number =	 1,
  pages =	 15,
  year =	 2024,
  publisher =	 {IOP Publishing}
}

@article{Iess-Stevenson-Parisi-et-al-2014:gravity,
  author =	 {Iess, L and Stevenson, D J and Parisi, M and
                  Hemingway, D and Jacobson, R A and Lunine, J I and
                  Nimmo, F and Armstrong, J W and Asmar, S W and
                  Ducci, M and Tortora, P},
  title =	 {{The Gravity Field and Interior Structure of
                  Enceladus}},
  journal =	 {Science},
  year =	 2014,
  volume =	 344,
  number =	 6179,
  pages =	 {78--80},
  month =	 apr
}

@article{Ivey-Imberger-1991:nature,
  title =	 {On the nature of turbulence in a stratified fluid.
                  Part I: The energetics of mixing},
  author =	 {Ivey, GN and Imberger, J},
  journal =	 {Journal of Physical Oceanography},
  volume =	 21,
  number =	 5,
  pages =	 {650--658},
  year =	 1991
}

@article{Jansen-Kang-Kite-et-al-2023:energetic,
  title =	 {Energetic constraints on ocean circulations of icy
                  ocean worlds},
  author =	 {Jansen, Malte F and Kang, Wanying and Kite, Edwin S
                  and Zeng, Yaoxuan},
  journal =	 {The Planetary Science Journal},
  volume =	 4,
  number =	 6,
  pages =	 117,
  year =	 2023,
  publisher =	 {IOP Publishing}
}

@article{Kang-2022:different,
  doi =		 {10.3847/1538-4357/ac779c},
  year =	 2022,
  publisher =	 {American Astronomical Society},
  volume =	 934,
  number =	 2,
  pages =	 116,
  author =	 {Wanying Kang},
  title =	 {Different Ice-shell Geometries on Europa and
                  Enceladus due to Their Different Sizes: Impacts of
                  Ocean Heat Transport},
  journal =	 {ApJ}
}

@article{Kang-2023:modulation,
  title =	 {The modulation effect of ice thickness variations on
                  convection in icy ocean worlds},
  author =	 {Kang, Wanying},
  journal =	 {Monthly Notices of the Royal Astronomical Society},
  volume =	 525,
  number =	 4,
  pages =	 {5251--5261},
  year =	 2023,
  publisher =	 {Oxford University Press}
}

@article{Kang-Bire-Marshall-2022:role,
  title =	 {The role of ocean circulation in driving hemispheric
                  symmetry breaking of the ice shell of Enceladus},
  journal =	 {EPSL},
  volume =	 599,
  pages =	 117845,
  year =	 2022,
  issn =	 {0012-821X},
  doi =		 {10.1016/j.epsl.2022.117845},
  author =	 {Wanying Kang and Suyash Bire and John Marshall}
}

@article{Kang-Jansen-2022:icy,
  title =	 {On Icy Ocean Worlds, Size Controls Ice Shell
                  Geometry},
  author =	 {Kang, Wanying and Jansen, Malte},
  journal =	 {ApJ},
  volume =	 935,
  number =	 2,
  pages =	 103,
  year =	 2022,
  publisher =	 {IOP Publishing}
}

@article{Kang-Mittal-Bire-et-al-2022:how,
  title =	 {How does salinity shape ocean circulation and ice
                  geometry on Enceladus and other icy satellites?},
  author =	 {Kang, Wanying and Mittal, Tushar and Bire, Suyash
                  and Campin, Jean-Michel and Marshall, John},
  journal =	 {Science Advances},
  volume =	 8,
  number =	 29,
  pages =	 {eabm4665},
  year =	 2022,
  publisher =	 {American Association for the Advancement of Science}
}

@article{Khurana-Kivelson-Stevenson-et-al-1998:induced,
  title =	 {Induced magnetic fields as evidence for subsurface
                  oceans in Europa and Callisto},
  author =	 {Khurana, KK and Kivelson, MG and Stevenson, DJ and
                  Schubert, G and Russell, CT and Walker, RJ and
                  Polanskey, C},
  journal =	 {Nature},
  volume =	 395,
  number =	 6704,
  pages =	 {777--780},
  year =	 1998,
  publisher =	 {Nature Publishing Group}
}

@article{Kivelson-Khurana-Russell-et-al-2000:galileo,
  title =	 {Galileo magnetometer measurements: A stronger case
                  for a subsurface ocean at Europa},
  author =	 {Kivelson, Margaret G and Khurana, Krishan K and
                  Russell, Christopher T and Volwerk, Martin and
                  Walker, Raymond J and Zimmer, Christophe},
  journal =	 {Science},
  volume =	 289,
  number =	 5483,
  pages =	 {1340--1343},
  year =	 2000,
  publisher =	 {American Association for the Advancement of Science}
}

@article{Kivelson-Khurana-Stevenson-et-al-1999:europa,
  title =	 {Europa and Callisto: Induced or intrinsic fields in
                  a periodically varying plasma environment},
  author =	 {Kivelson, MG and Khurana, KK and Stevenson, DJ and
                  Bennett, L and Joy, S and Russell, CT and Walker, RJ
                  and Zimmer, C and Polanskey, C},
  journal =	 {JGR: Space Physics},
  volume =	 104,
  number =	 {A3},
  pages =	 {4609--4625},
  year =	 1999,
  publisher =	 {Wiley Online Library}
}

@article{Lecoanet-Quataert-2013:internal,
  title =	 {Internal gravity wave excitation by turbulent
                  convection},
  author =	 {Lecoanet, Daniel and Quataert, Eliot},
  journal =	 {Monthly Notices of the Royal Astronomical Society},
  volume =	 430,
  number =	 3,
  pages =	 {2363--2376},
  year =	 2013,
  publisher =	 {Oxford University Press}
}

@article{Liao-Nimmo-Neufeld-2020:heat,
  title =	 {Heat production and tidally driven fluid flow in the
                  permeable core of Enceladus},
  author =	 {Liao, Yang and Nimmo, Francis and Neufeld, Jerome A},
  journal =	 {JGR: Planets},
  volume =	 125,
  number =	 9,
  pages =	 {e2019JE006209},
  year =	 2020,
  publisher =	 {Wiley Online Library}
}

@Article{Marshall-Adcroft-Hill-et-al-1997:finite,
  author =	 {J. Marshall and A. Adcroft and C. Hill and L.
                  Perelman and C. Heisey},
  year =	 1997,
  title =	 {A finite-volume, incompressible {Navier Stokes}
                  model for studies of the ocean on parallel
                  computers},
  journal =	 JGR,
  volume =	 102,
  pages =	 {5,753--5,766},
}

@article{McCarthy-Cooper-2016:tidal,
  author =	 {McCarthy, Christine and Cooper, Reid F},
  title =	 {{Tidal dissipation in creeping ice and the thermal
                  evolution of Europa}},
  journal =	 {Earth and Planetary Science Letters},
  year =	 2016,
  volume =	 443,
  pages =	 {185--194}
}

@article{McDougall-Barker-2011:getting,
  title =	 {Getting started with TEOS-10 and the Gibbs Seawater
                  (GSW) oceanographic toolbox},
  author =	 {McDougall, Trevor J and Barker, Paul M},
  journal =	 {SCOR/IAPSO WG},
  volume =	 127,
  pages =	 {1--28},
  year =	 2011
}

@article{McDougall-Jackett-Wright-et-al-2003:accurate,
  author =	 {McDougall, Trevor J and Jackett, David R and Wright,
                  Daniel G and Feistel, Rainer},
  title =	 {{Accurate and Computationally Efficient Algorithms
                  for Potential Temperature and Density of Seawater}},
  journal =	 {Journal of Atmospheric and Oceanic Technology},
  year =	 2003,
  volume =	 20,
  number =	 5,
  pages =	 {730--741},
  month =	 may
}

@article{Melosh-Ekholm-Showman-et-al-2004:temperature,
  title =	 {The temperature of Europa's subsurface water ocean},
  author =	 {Melosh, HJ and Ekholm, AG and Showman, AP and
                  Lorenz, RD},
  journal =	 {Icarus},
  volume =	 168,
  number =	 2,
  pages =	 {498--502},
  year =	 2004,
  publisher =	 {Elsevier}
}

@article{Munk-Wunsch-1998:abyssal,
  author =	 {W. Munk and C. Wunsch},
  title =	 {Abyssal recipes II: energetics of tidal and wind
                  mixing},
  journal =	 {Deep-sea Research Part I-oceanographic Research
                  Papers},
  volume =	 45,
  number =	 12,
  pages =	 {1977-2010},
  month =	 dec,
  year =	 1998
}

@article{Nimmo-Barr-Behounkova-et-al-2018:thermal,
  title =	 {The thermal and orbital evolution of Enceladus:
                  observational constraints and models},
  author =	 {Nimmo, Francis and Barr, Amy C and Behounkov{\'a},
                  Marie and McKinnon, William B},
  journal =	 {Enceladus and the icy moons of Saturn},
  volume =	 475,
  pages =	 {79--94},
  year =	 2018,
  publisher =	 {University of Arizona Press Tucson}
}

@article{Osborn-1980:estimates,
  title =	 {Estimates of the local rate of vertical diffusion
                  from dissipation measurements},
  author =	 {Osborn, Thomas R},
  journal =	 {Journal of physical oceanography},
  volume =	 10,
  number =	 1,
  pages =	 {83--89},
  year =	 1980
}

@article{Pappalardo-Belton-Breneman-et-al-1999:does,
  title =	 {Does Europa have a subsurface ocean? Evaluation of
                  the geological evidence},
  author =	 {Pappalardo, Robert T and Belton, Michael JS and
                  Breneman, HH and Carr, MH and Chapman, Clark R and
                  Collins, GC and Denk, T and Fagents, S and Geissler,
                  Paul E and Giese, B and others},
  journal =	 {JGR: Planets},
  volume =	 104,
  number =	 {E10},
  pages =	 {24015--24055},
  year =	 1999,
  publisher =	 {Wiley Online Library}
}

@article{Park-Mastrodemos-Jacobson-et-al-2024:global,
  title =	 {The global shape, gravity field, and libration of
                  Enceladus},
  author =	 {Park, RS and Mastrodemos, N and Jacobson, RA and
                  Berne, A and Vaughan, AT and Hemingway, DJ and
                  Leonard, EJ and Castillo-Rogez, JC and Cockell, CS
                  and Keane, JT and others},
  journal =	 {Journal of Geophysical Research: Planets},
  volume =	 129,
  number =	 1,
  pages =	 {e2023JE008054},
  year =	 2024,
  publisher =	 {Wiley Online Library}
}

@article{Peltier-Caulfield-2003:mixing,
  title =	 {Mixing efficiency in stratified shear flows},
  author =	 {Peltier, WR and Caulfield, CP},
  journal =	 {Annual review of fluid mechanics},
  volume =	 35,
  number =	 1,
  pages =	 {135--167},
  year =	 2003
}

@article{Ramadhan-Wagner-Hill-et-al-2020:oceananigans,
  doi =		 {10.21105/joss.02018},
  year =	 2020,
  publisher =	 {The Open Journal},
  volume =	 5,
  number =	 53,
  pages =	 2018,
  author =	 {Ali Ramadhan and Gregory LeClaire Wagner and Chris
                  Hill and Jean-Michel Campin and Valentin Churavy and
                  Tim Besard and Andre Souza and Alan Edelman and
                  Raffaele Ferrari and John Marshall},
  title =	 {Oceananigans.jl: Fast and friendly geophysical fluid
                  dynamics on GPUs},
  journal =	 {Journal of Open Source Software}
}

@article{Rekier-Trinh-Triana-et-al-2019:internal,
  title =	 {Internal Energy Dissipation in Enceladus's
                  Subsurface Ocean From Tides and Libration and the
                  Role of Inertial Waves},
  author =	 {Rekier, Jeremy and Trinh, Antony and Triana, SA and
                  Dehant, V{\'e}ronique},
  journal =	 {Journal of Geophysical Research: Planets},
  volume =	 124,
  number =	 8,
  pages =	 {2198--2212},
  year =	 2019,
  publisher =	 {Wiley Online Library}
}

@article{Roberts-2015:fluffy,
  author =	 {Roberts, James H},
  title =	 {{The fluffy core of Enceladus}},
  journal =	 {Icarus},
  year =	 2015,
  volume =	 258,
  pages =	 {54--66},
  month =	 sep
}

@article{Robuchon-Choblet-Tobie-et-al-2010:coupling,
  title =	 {Coupling of thermal evolution and despinning of
                  early Iapetus},
  author =	 {Robuchon, G and Choblet, G and Tobie, G and
                  {\v{C}}adek, O and Sotin, C and Grasset, O},
  journal =	 {Icarus},
  volume =	 207,
  number =	 2,
  pages =	 {959--971},
  year =	 2010,
  publisher =	 {Elsevier}
}

@article{Rovira-Navarro-Katz-Liao-et-al-2022:tides,
  title =	 {The Tides of Enceladus’ Porous Core},
  author =	 {Rovira-Navarro, Marc and Katz, Richard F and Liao,
                  Yang and van der Wal, Wouter and Nimmo, Francis},
  journal =	 {JGR: Planets},
  pages =	 {e2021JE007117},
  year =	 2022,
  publisher =	 {Wiley Online Library}
}

@article{Rovira-Navarro-Rieutord-Gerkema-et-al-2019:do,
  title =	 {Do tidally-generated inertial waves heat the
                  subsurface oceans of Europa and Enceladus?},
  author =	 {Rovira-Navarro, Marc and Rieutord, Michel and
                  Gerkema, Theo and Maas, Leo RM and van der Wal,
                  Wouter and Vermeersen, Bert},
  journal =	 {Icarus},
  volume =	 321,
  pages =	 {126--140},
  year =	 2019,
  publisher =	 {Elsevier}
}

@article{Russell-Murray-Hand-2017:possible,
  title =	 {The possible emergence of life and differentiation
                  of a shallow biosphere on irradiated icy worlds: the
                  example of Europa},
  author =	 {Russell, Michael J and Murray, Alison E and Hand,
                  Kevin P},
  journal =	 {Astrobiology},
  volume =	 17,
  number =	 12,
  pages =	 {1265--1273},
  year =	 2017,
  publisher =	 {Mary Ann Liebert, Inc. 140 Huguenot Street, 3rd
                  Floor New Rochelle, NY 10801 USA}
}

@article{Schenk-McKinnon-2024:new,
  title =	 {New global topography of Enceladus: Hypsometry,
                  basins, spherical harmonics, shell thickness, and
                  true polar wander revisited},
  author =	 {Schenk, Paul M and McKinnon, William B},
  journal =	 {Icarus},
  volume =	 408,
  pages =	 115827,
  year =	 2024,
  publisher =	 {Elsevier}
}

@article{Shoji-Hussmann-Kurita-et-al-2013:ice,
  title =	 {Ice rheology and tidal heating of Enceladus},
  author =	 {Shoji, Daigo and Hussmann, H and Kurita, K and Sohl,
                  F},
  journal =	 {Icarus},
  volume =	 226,
  number =	 1,
  pages =	 {10--19},
  year =	 2013,
  publisher =	 {Elsevier}
}

@Article{Soderlund-Schmidt-Wicht-et-al-2014:ocean,
  author =	 {Soderlund, K. M. and Schmidt, B. E. and Wicht, J.
                  and Blankenship, D. D.},
  title =	 {Ocean-driven heating of Europa's icy shell at low
                  latitudes},
  journal =	 {Nature Geoscience},
  year =	 2014,
  volume =	 7,
  pages =	 {16-19}
}

@article{Soucek-Behounkova-Cadek-et-al-2019:tidal,
  author =	 {Soucek, Ondrej and Behounkova, Marie and Cadek,
                  Ondrej and Hron, Jaroslav and Tobie, Gabriel and
                  Choblet, Gael},
  title =	 {{Tidal dissipation in Enceladus' uneven, fractured
                  ice shell}},
  journal =	 {Icarus},
  year =	 2019,
  volume =	 328,
  pages =	 {218--231},
  month =	 aug
}

@article{Stommel-1961:thermohaline,
  author =	 {Stommel, H.},
  title =	 {Thermohaline convection with two stable regimes of
                  flow},
  journal =	 {Tellus},
  year =	 1961,
  volume =	 13,
  pages =	 {224-230},
}

@article{Taubner-Pappenreiter-Zwicker-et-al-2018:biological,
  title =	 {Biological methane production under putative
                  Enceladus-like conditions},
  author =	 {Taubner, Ruth-Sophie and Pappenreiter, Patricia and
                  Zwicker, Jennifer and Smrzka, Daniel and Pruckner,
                  Christian and Kolar, Philipp and Bernacchi,
                  S{\'e}bastien and Seifert, Arne H and Krajete,
                  Alexander and Bach, Wolfgang and others},
  journal =	 {Nature comm.},
  volume =	 9,
  number =	 1,
  pages =	 {1--11},
  year =	 2018,
  publisher =	 {Nature Publishing Group}
}

@article{Thomas-Tajeddine-Tiscareno-et-al-2016:enceladus,
  title =	 {Enceladus’s measured physical libration requires a
                  global subsurface ocean},
  author =	 {Thomas, PC and Tajeddine, R and Tiscareno, MS and
                  Burns, JA and Joseph, J and Loredo, TJ and
                  Helfenstein, P and Porco, C},
  journal =	 {Icarus},
  volume =	 264,
  pages =	 {37--47},
  year =	 2016,
  publisher =	 {Elsevier}
}

@article{Tyler-2020:heating,
  title =	 {Heating of Enceladus due to the dissipation of ocean
                  tides},
  author =	 {Tyler, Robert H},
  journal =	 {Icarus},
  volume =	 348,
  pages =	 113821,
  year =	 2020,
  publisher =	 {Elsevier}
}

@Book{Vallis-2006:atmospheric,
  author =	 {Vallis, G. K.},
  title =	 {Atmospheric and oceanic fluid dynamics, fundamentals
                  and large-scale circulation},
  publisher =	 {Cambridge University Press},
  year =	 2006
}

@article{Wunsch-Ferrari-2004:vertical,
  author =	 {C. Wunsch and R. Ferrari},
  title =	 {Vertical mixing, energy and thegeneral circulation
                  of the oceans},
  journal =	 ARFM,
  volume =	 36,
  pages =	 {281-314},
  year =	 2004
}

@article{Zeng-Jansen-2021:ocean,
  title =	 {Ocean circulation on enceladus with a high-versus
                  low-salinity ocean},
  author =	 {Zeng, Yaoxuan and Jansen, Malte F},
  journal =	 {PSJ},
  volume =	 2,
  number =	 4,
  pages =	 151,
  year =	 2021,
  publisher =	 {IOP Publishing}
}

@article{Zeng-Jansen-2024:effect,
  title =	 {The Effect of Salinity on Ocean Circulation and
                  Ice--Ocean Interaction on Enceladus},
  author =	 {Zeng, Yaoxuan and Jansen, Malte F},
  journal =	 {The Planetary Science Journal},
  volume =	 5,
  number =	 1,
  pages =	 13,
  year =	 2024,
  publisher =	 {IOP Publishing}
}

@article{Zhang-Kang-Marshall-2024:ocean,
  author =	 {Yixiao Zhang and Wanying Kang and John Marshall },
  title =	 {Ocean weather systems on icy moons, with application
                  to Enceladus},
  journal =	 {Science Advances},
  volume =	 10,
  number =	 45,
  pages =	 {eadn6857},
  year =	 2024,
  doi =		 {10.1126/sciadv.adn6857}
}

@misc{Zhang-Kang-Marshall-2025:how,
  title =	 {How does ice shell geometry shape ocean dynamics on
                  icy moons?},
  author =	 {Yixiao Zhang and Wanying Kang and John Marshall},
  year =	 2025,
  eprint =	 {2510.25988},
  archivePrefix ={arXiv},
  primaryClass = {astro-ph.EP},
  url =		 {https://arxiv.org/abs/2510.25988},
}

@article{Zhu-Manucharyan-Thompson-et-al-2017:influence,
  title =	 {The influence of meridional ice transport on
                  Europa's ocean stratification and heat content},
  author =	 {Zhu, Peiyun and Manucharyan, Georgy E and Thompson,
                  Andrew F and Goodman, Jason C and Vance, Steven D},
  journal =	 {Geophysical Research Letters},
  volume =	 44,
  number =	 12,
  pages =	 {5969--5977},
  year =	 2017,
  publisher =	 {Wiley Online Library}
}

@article{Zimmer-Khurana-Kivelson-2000:subsurface,
  title =	 {Subsurface oceans on Europa and Callisto:
                  Constraints from Galileo magnetometer observations},
  author =	 {Zimmer, Christophe and Khurana, Krishan K and
                  Kivelson, Margaret G},
  journal =	 {Icarus},
  volume =	 147,
  number =	 2,
  pages =	 {329--347},
  year =	 2000,
  publisher =	 {Elsevier}
}

%
%
%
%
%

\end{document}